\documentclass[a4paper]{revtex4}
\usepackage[english]{babel}
\usepackage{array,booktabs}
\usepackage{array} 
\usepackage{lipsum}   
\usepackage{calc}
\usepackage{pdflscape}
\usepackage{color}
\usepackage{float}
\usepackage[font=small,labelfont=bf]{caption}
\usepackage{graphicx}
\usepackage{amsmath}
\usepackage[nodisplayskipstretch]{setspace}
\renewcommand{\arraystretch}{1.9}
\usepackage{setspace}
\usepackage{tabularx}

\usepackage{float}
\usepackage{color}
\usepackage{amsmath}
\usepackage{float}
\usepackage{calc}
\usepackage{pdflscape}
\usepackage{color}
\usepackage{float}
\usepackage{subfigure}
\usepackage[font=small,labelfont=bf]{caption}
\usepackage{graphicx}
\begin{document}{\setlength\abovedisplayskip{4pt}}

\title{Significance of broken $ \mu-\tau $ Symmetry in correlating $ \delta_{CP} $, $ \theta_{13} $, Lightest neutrino Mass and neutrinoless double beta decay $ 0\nu\beta\beta $}

\author{Gayatri Ghosh}
\email{gayatrighdh@gmail.com}
\affiliation{Department of Physics, Gauhati University, Jalukbari, Assam-781015, India}
\affiliation{Department of Physics, Pandit Deendayal Upadhayay Mahavidyalaya, Karimganj, Assam-788720, India}

\begin{abstract}
Leptonic CP Violating Phase $ \delta_{CP} $ in the light neutrino sector and leptogenesis via present matter antimatter asymmetry of the Universe entails each other. Probing CP violation in light neutrino oscillation is one of the challenging tasks today. The reactor mixing angle $ \theta_{13} $ measured in reactor experiments, LBL, DUNE with high precision in neutrino experiments indicates towards the vast dimension of scope to detect $ \delta_{CP} $. The correlation between leptonic Dirac CPV phase $ \delta_{CP} $, reactor mixing angle $ \theta_{13} $, lightest neutrino mass $ m_{1} $ and matter antimatter asymmetry of the Universe within the framework of $ \mu-\tau $ symmetry breaking assuming the type I seesaw dominance is extensively studied here. Small tiny breaking of the $ \mu-\tau $ symmetry allows a large Dirac CP violating phase in neutrino oscillation which in turn is characterised by awareness of measured value of $ \theta_{13} $ and to provide a hint towards a better understanding of the experimentally observed near maximal value of $ \nu_{\mu} -\nu_{\tau} $ mixing angle $ \theta_{23}\simeq \frac{\pi}{4}$. Precise breaking of the $ \mu-\tau $ symmetry is achieved by adding a 120 plet Higgs to the 10 $+$ $\bar{126}$ dimensional representation of Higgs. The estimated three dimensional density parameter space of lightest neutrino mass $ m_{1} $, $ \delta_{CP} $, reactor mixing angle $ \theta_{13} $, is constrained here for the requirement of producing the observed value of baryon asymmetry of the Universe through the mechanism of leptogenesis. Carrying out numerical analysis the allowed parameter space of  $ m_{1} $, $ \delta_{CP} $, $ \theta_{13} $, is found out which can produce the observed baryon to photon density ratio of the Universe.  
\end{abstract}   
\maketitle
\section{Introduction}
\label{intro}
In 1950, Brune Pontecorvo for the first time emphasized the idea of neutrino oscillations which resembled $ K^{0}-\bar{K^{0}} $ oscillations. In neutrino oscillations, a neutrino originated with a definite flavor, ($ \nu_{e}, \nu_{\mu}, \nu_{\tau} $) oscillates to a distinct contrasting lepton flavor. neutrino oscillation reveals that each of the three states of neutrino $ \nu_{\alpha} $ in flavor basis is a superposition of three mass eigen states ($ m_{1},m_{2},m_{3} $) {\color{blue}\cite{1}}. neutrinos are massive and they mix with each other. The massive neutrinos are formed in their gauge eigen states $(\nu_{\alpha})$ which is linked to their mass eigen states $ \nu_{i} $. Gauge eigen states participate in gauge interactions as 
\begin{equation}
\mid \nu_{\alpha} > = \sum U_{\alpha_{i}}\mid \nu_{i} >
\end{equation} 
where, $ \alpha = e, \mu, \tau$ , $ \nu_{i} $ is the neutrino of distinct mass $ m_{i} $.U is parameterised as  
\begin{equation}
U = \begin{pmatrix}
c_{12}c_{13} & s_{12}c_{13} & s_{13}e^{-i\delta}\\
-s_{12}c_{23}-c_{12}s_{23}s_{13}e^{i\delta} & c_{12}c_{23}-s_{12}s_{23}s_{13}e_{i\delta}& s_{23}c_{13}\\
s_{12}s_{23}-c_{12}c_{23}s_{13}e^{i\delta} & -c_{12}s_{23}-s_{12}c_{23}s_{13}e^{i\delta} & c_{23}c_{13}\\
\end{pmatrix}
\end{equation}
where, $\theta_{12} = 33^{0}, \theta_{23} = 38^{0}-53^{0}, \theta_{13} = 8^{0}$ {\color{blue}\cite{2}} are the solar, atmospheric and reactor angles according to the global fits respectively. The Majorana phases $ \alpha $, $ \beta $ dwells in P, where 
\begin{equation}
P = \text{diag} \begin{pmatrix} 1 & e^{i\alpha} & e^{i(\beta + \delta)}\end{pmatrix}
\end{equation}
$ U*P $ is known as the Pontecorvo-Maki-Nakagawa-Sakata $U_{PMNS}$ matrix {\color{blue}\cite{3}}. Since a $ \nu $ of a given flavor $ \alpha $ is a mixed state of atleast three $ \nu $ with distinct masses, this three generation mixing could result into the flavor mixing mass matrix or PMNS matrix possessing an irreducible imaginary component. This irreducible imaginary component is responsible for CP asymmetry. 
CP violation interchanges every particle into its antiparticle. $ \delta_{CP} $ in PMNS matrix can induce CP violation. CP asymmetry can be observed in neutrino oscillations. $ \delta_{CP} $ phase measures the amount of asymmetries between lepton oscillations and antilepton oscillations. neutrinos being massive and they mix with each other. This may be a source of CP violation if Sin$ \delta_{CP} \neq 0$. The amount of $ \delta_{CP} $ violation phase in this case is estimated by the Jarkslog invariant {\color{blue}\cite{4}}.
\begin{equation}
J_{CP} =\frac{1}{8}Cos\theta_{13}Sin2\theta_{12}Sin2\theta_{23}Sin2\theta_{13}Sin\delta_{CP}
\end{equation}
when $ Sin \delta_{CP} \neq 0$
In leptogenesis, lepton-antilepton asymmetry is explained if there are complex imaginary irreducible terms in the yukawa couplings of lepton mass matrices. The lepton number generation of the early Universe can be estimated by the complex CPV phase term in the fermion mass matrices. The ongoing T2K experiment have reported that CP violating phase, $ \delta_{CP} $ excludes the value $ \delta_{CP} = 0,\pi $ {\color{blue}\cite{5}} at the 2$ \sigma $ confidence interval for either of the mass orderings, Normal ordering and Inverted ordering. The value of Dirac CPV phase, $ \delta_{CP} = 276.5^{0}$ is preferred in {\color{blue}\cite{6}}.
The neutrino mass matrix is invariant under $ \mu-\tau $ exchange symmetry. In a basis where the charged leptons are mass eigen states. Under the $ \mu -\tau $ exchange symmetry, the 2-3 mixing is maximal, i.e, $ \theta_{23} = \frac{\pi}{4}$ and the 1-3 mixing is zero, i.e, $ \theta_{13}=0 $.
The deviation of $ \delta\theta_{23} $ from the maximal angle $ \theta_{23} $, the explanation of reactor angle $ \theta_{13} $ and the existence of CP violating phase neccessitates of the spontaneous breaking of the $ \mu-\tau $ exchange symmetry in the neutrino sector.
Here, in this work, an explicit form of the Dirac neutrino mass matrix in broken $ \mu-\tau $ {\color{blue}\cite{7}} symmetry framework in type I Seesaw mechanism is used in our calculation for generating baryon asymmetry of the Universe via leptogenesis. This scenario is characterised by small divergence of $ \delta \theta_{23} $ from the maximal angle $ \theta_{23} $, which is consistent with a liberal size of $ \theta_{13}\sim 8^{0}-9^{0} $ and a large $ \delta_{CP} $ phase in the neutrino sector. The renormalisable Dirac neutrino yukawa couplings of the Dirac mass matrices is determined from the fermion yukawa couplings to the 10, $ \bar{126} $ and 120 dimensional fields of Higgs multiplet in the SO(10) group. Higgs field under the 10 and $ \bar{126} $ representations are symmetrical under the generalised $ \mu-\tau $ symmetry, while the 120 dimensional representation changes sign. This spontaneously breaks the $ \mu-\tau $ invariant symmetry, which in turn allows a generalised $ \delta_{CP} $ phase in the PMNS matrix.
Here we made an effort for correlating or constraining the values of $ \delta_{CP} $ phase, non zero reactor angle $ \theta_{13} $ and lightest neutrino mass space for both the hierarchies in the context of leptogenesis and current ratio of baryon to photon density of the Universe.
Both CPV phase $ \delta_{CP} $ and reactor angle $ \theta_{13} $ have good vibes between each other. Precise value of $ \theta_{13} $ plays an imperative role in its CP violation phase measurements. On the basis of this fact non zero values of $ \theta_{13} $ is predicted here in consistency with $ \delta_{CP} $ phase. Taking into account constraints from the global fit values of $ \nu $ oscillation parameters and cosmology a density plot of the favourable values of $ \delta_{CP} $ phase, lightest neutrino mass, $ \theta_{13} $ is being initiated, which is compatible with the contraints on the sum of the absolute neutrino masses, $ \sum_{i} m(\nu_{i)} < 0.23 \hspace{.1cm}\text{eV}$ from CMB, Planck 2015 data (CMB15+ LRG+ lensing + $H_{0}$) {\color{blue}\cite{8}}. Constraints fom the leptonic asymmetry of the Universe is also considered for further restricting $ \delta_{CPV} $ phase space and lightest neutrino mass. The leptonic CP asymmetry is being deliberated via leptogenesis in terms of baryon density to photon density ratio of the Universe $ \eta_{B} $ accessible as $5.8 \times 10^{-10} <\eta_{B}<6.6 \times 10^{-10}$ {\color{blue}\cite{9}}.
We also calculate the effective mass spectrum for neutrinoless double beta decay, $0\nu\beta\beta$ decay given by
\begin{equation}
m_{ee} = |m_{1}c^{2}_{12}c^{2}_{13} + m_{2}s^{2}_{12}c^{2}_{13}e^{i\alpha_{21}} +m_{3}s^{2}_{13}e^{i(\alpha_{31}-2\delta_{CP})}|
\end{equation}
for favourable values of $ \delta_{CP} $ phase and lightest $ \nu $ mass explored here in this work. 
In this paper, we apply the broken $ \mu-\tau $ symmetry to the Dirac neutrino Yukawa Couplings in type-I seesaw mechanism in SO(10) model in predicting favourable values of $ \delta_{CP} $ phase, lightest neutrino mass and $ \theta_{13} $. We then scan free parameters in these models and search for the allowed region in which the neutrino oscillation data can be fitted. For the allowed parameter sets, we show the predictions of observables like $ \delta_{CP} $ phase, lightest neutrino mass and $ \theta_{13} $ in
the neutrino sector. Finally we show our predictions for the the effective mass spectrum for neutrinoless double beta, $0\nu\beta\beta$ decay for favourable values of $ \delta_{CP} $ phase.
\par
The paper is organized as follows. In section 2, we introduce our broken $\mu-\tau $ symmetry models. In section 3, we perform parameter scan to fit neutrino oscillation data and provide some informations in predicting observables like $ \delta_{CP} $ phase in the neutrino sector. Also  we show our predictions for the the effective mass spectrum for neutrinoless double beta, $0\nu\beta\beta$ decay for favourable values of $ \delta_{CP} $ phase. Section 4 is our results and conclusions for the  Yukawa interactions associated with broken $ \mu-\tau $ symmetry model discussed here.
\section{Broken $\mu-\tau$ symmetry with type I Seesaw Mechanism}
The Lagrangian of the type I Seesaw model is {\color{blue}\cite{10}}
\begin{equation}
L_{N}^{M} = \frac{i}{2}\bar{N^{R}_{I}}\delta N_{I}^{R} - y_{I\alpha}\bar{N}_{I}^{R}\tilde{\phi^{\dagger}}L_{\alpha} - \frac{1}{2}\bar{N^{R}_{I}}M_{IJ}(N_{J}^{R})^{C} + h.c
\end{equation}
Here, $y_{I\alpha}$ is the complex yukawa coupling matrix,$ L_{\alpha} = (\nu_{\alpha}^{L}, L_{\alpha}^{l})^{T}  $ is the standard model left handed lepton doublet of flavor $ \alpha $, when $ \alpha = e, \mu, \tau $ and $\tilde{\phi}$ is the hypercharge conjugated Higgs Doublet, $ (\phi^{*}_{0}, -\phi_{-})^{T} $. 
\par 
The Lagrangian describes scenario of generation of $ \nu  $ masses via Higgs mechanism. Electroweak symmetry breaking process allows neutral part of the Higgs field to acquire a VEV, $ v = 246 GeV $, $ \sqrt{2} <\phi_{0}> = v$ so that the left handed and right handed neutrinos forms massive Dirac fermions.
\par 
In Eq. (6), $M_{IJ}$ is a symmetric matrix of left handed violating Majorana masses, where $ M_{IJ} = M_{I}\delta_{IJ} $ is real and diagonal. Here right handed neutrino masses is larger than the Electroweak Scale. The $ \nu $ masses is then suppresed by right handed neutrino yukawa couplings, and also by
\begin{equation}
\frac{v}{M_{I}}< 1
\end{equation}
In type I Seesaw, the baryon asymmetry of the Universe (BAU) occurs via leptogenesis mechanism via out of equilibrium decay of heavy RH Majorana neutrinos in the early Universe via electroweak sphaleron processes {\color{blue}\cite{11}}. The resulting Majorana mass matrix of light SM neutrinos is
\begin{equation}
M_{\alpha \beta} = -(m_{D}^{T})_{\alpha I}(M_{IJ}^{-1})(m_{D})_{J \beta}
\end{equation}
where $(m_{D})_{\alpha I}$ is the Dirac mass matrix. Eq. 8 shows that in type I seesaw mechanism, SM $ \nu $ masses are suppressed by the combination of small yukawa couplings and large RH $ \nu $ masses. neutrino mass matrix on diagonalision, gives two eigen values $ - $ light neutrino $ \sim $ $ \frac{m^{2}_{D}}{M_{R}} $ and a heavy neutrino state $ \sim $ $M_{R}$. This is known as type I seesaw mechanism. 
\par 
In SO(10), handed heavy Majorana neutrino couples to the left handed $ \nu $ via Dirac mass matrix $m_{D}$. Out of the decay of the lightest of the  RH Majorana neutrinos, $M_{1}$, i.e $M_{3}, M_{2}\gg M_{1}$ will contribute to CP asymmetry \cite{6,12} (for leptogenesis), i.e $\epsilon^{CP}_{l}$ and leptogenesis \cite{SO1,SO2,SO3,SO4,SO5,SO6,SOF}.
\par 
At the end of inflation {\color{blue}\cite{13}}, a certain number density of right-handed neutrinos, $ n_{\nu_{R}} $, were created, which is linked to the present cosmological scenario. Right-handed neutrinos decayed, with a decay rate that reads, at tree level 
\begin{equation}
\Gamma_{D_{i}} = \Gamma(\nu_{R_{i} \rightarrow l_{i} + H_{u} } + \Gamma(\nu_{R_{i} \rightarrow \tilde{l_{i}} + \tilde{H_{u}} }) = \frac{1}{8}(Y_{\nu}Y_{\nu}^{\dagger})_{ii}M_{i}
\end{equation}
It is convenient to work in a basis of right-handed
neutrinos where RH $\nu$ mass matrix is diagonal, the type I contribution to $\epsilon^{CP}_{l}$ is given by decay of $M_{1}$ or the CP violating parameter is given as,
 \begin{equation}
\epsilon^{CP}_{l}=\frac{\Gamma_{D_{i}}-\bar{\Gamma_{D_{i}}}}{\Gamma_{D_{i}}+\bar{\Gamma_{D_{i}}}}, 
\end{equation}
where
\begin{equation}
\Gamma_{D_{i}} = \Gamma(\nu_{R_{i}}\rightarrow l_{i}H_{u}) + \Gamma(\nu_{R_{i}}\rightarrow \tilde{L_{i}}\tilde{h_{u}}) = \frac{1}{8}(Y_{\nu}Y_{\nu}^{\dagger})M_{i}
\end{equation}
where $\Gamma(\nu_{R_{i}}\rightarrow l_{i}H_{u})$ means decay rate of heavy Majorana RH $\nu$ of mass $M_{1}$ to a lepton and Higgs. In electroweak sphaleron process,  asymmetries produced by the out of equilibrium decay of $M_{2}$ and $ M_{3} $ gets washed out by lepton number violating interactions after $ \nu_{R} $ or $ M_{1} $ decay. In lepton number violating interactions (decays, inverse decays and scatterings) must be out of equilibrium when the right-handed neutrinos decay. In the basis where RH $\nu$ mass matrix is diagonal, the type I contribution to $\epsilon^{CP}_{l}$ is given by \cite{Mh},
\begin{equation}
\epsilon_{CP}=-\frac{3M_{1}}{8\pi}\frac{Im[\Delta m^{2}_{\odot}Q^{2}_{12}+\Delta m^{2}_{A}Q^{2}_{13}]}{\upsilon^{2}\sum
|Q_{1j}|^{2}m_{j}}. 
\end{equation}
 where $\upsilon$ is the Higg's vev. $Q$ is a complex unitary orthogonal matrix  where $Q$ is parameterized as {\color{blue}\cite{Osc}}:$ Q = D_{\sqrt{M^{-1}}}Y_{\nu}UD_{\sqrt{K^{-1}}} $ where $Y_{\nu}$ is the Dirac neutrino Yukawa couplings. To reproduce the physical, low-energy, parameters, i.e. the light neutrino masses (contained in $ D_{K} $) and mixing angles and CP phases (contained in $ U_{PMNS} $), we have taken the most general Dirac neutrino mass matrix in broken $\mu-\tau$ symmetry framework as {\color{blue}\cite{kod}}
 \begin{equation}
Y_{\nu} = \frac{M_{\nu}}{246 \hspace{0.2cm} GeV} = \frac{1}{246 \hspace{0.2cm} GeV}\begin{pmatrix}
11353.7 & -11193.7 + 12692.2 i & -11193.7 - 12692.2 i\\
-11193.7 - 12692.2 i & 62440.4 & 62279.4-14582.3 i\\
-11193.7 + 12692.2 i & 62279.4+14582.3 i & 62021.3\\
\end{pmatrix}
\end{equation}
 In the flavor basis, where the charged-lepton Yukawa matrix, $Y_{e}$ and gauge interactions are flavour-diagonal, 
\begin{equation}
D_{K} = U^{T}Y_{\nu}^{T}D_{M}^{-1}Y_{\nu}U
\end{equation} 
 In terms of user defined Dirac neutrino mass matrices, {\color{blue}\cite{kod}}
 \begin{equation}
  K=Y_{\nu}^{T}M_{R}^{-1}Y_{\nu}
\end{equation}  $U$ is the PMNS matrix and  $M_{R}$ is the RH neutrino Majorana scale. 
We can always choose to work in a basis of right neutrinos where
M is diagonal $D_{M} = Diag(M_{1}, M_{2},M_{3})$ where $M_{3}, M_{2}\gg M_{1}$. Equation (12) expresses $  \epsilon_{CP}$ in terms of both the solar ($\Delta m^{2}_{21}$) and atmospheric ($ \Delta m^{2}_{A} $) mass squared differences, eq.(12) also reveals that CP asymmetry is linked to Dirac CPV phase. Here we utilse this fact to generate the allowed region of $ \delta_{CP} $ phase in context of laptogenesis. As has been discussed in \cite{Mh}, the lepton $ –$ antilepton asymmetry gets connected to both the solar and the atmospheric mass difference square. The transformation of the lepton asymmetry into a baryon asymmetry by non-perturbative $B+L$ violating (sphaleron) processes is discussed in {\color{blue}\cite{6}}. 
\par 
neutrino masses and mixings are connected with the atmospheric and
solar neutrino fluxes, this is suitable to explain flavor changing neutral current processes, FCNC processes, like $\mu \rightarrow e + \gamma  $ processes. In supersymmetric theories like cMSSM, NUHM, NUGM and NUSM where the origin of the $ \nu $ masses is via see-saw mechanism, it can be shown that the prediction for $BR(\mu \rightarrow  e, \gamma)$, $BR(\tau \rightarrow  \mu, \gamma)$ and $BR(\tau \rightarrow  e, \gamma)$ is in general larger than the experimental upper MEG bound {\color{blue}\cite{GG,GG1}}. Also some studies on decays of b flavoured hadrons in the context of cMSSM/mSUGRA models is being done in {\color{blue}\cite{bfh}}.  
\par 
Let H, F (G) be any complex symmetric (anti) matrices in general, which are a measure of the fermion yukawa couplings to the 10, $\bar{126}$ and 120 Higgs field respectively. 
Here,
\begin{equation}
H = \begin{pmatrix}
h_{11} & h_{12} & h_{12}\\
h_{12} & h_{22} & h_{23}\\
h_{12} & h_{23} & h_{22}\\\\
\end{pmatrix}
\end{equation}
Similarly
\begin{equation}
F = \begin{pmatrix}
f_{11} & f_{12} & f_{12}\\
f_{12} & f_{22} & f_{23}\\
f_{12} & f_{23} & f_{22}\\\\
\end{pmatrix}
\end{equation}
Similarly
\begin{equation}
G = \begin{pmatrix}
0 & g_{12} & -g_{12}\\
-g_{12} & 0  & g_{23}\\
g_{12} & -g_{23} & 0\\\\
\end{pmatrix}
\end{equation}
\par
h,f,g are all real. $ \mu-\tau  $ symmetry is invariant under the exchange of second and third generation fermions. When $ \mu-\tau  $ symmetry is added with SO(10) grand unified theory then a general symmetry results which satisfies
\begin{equation}
S^{T}(H,F,G)S = (H,F,-G) 
\end{equation}
where
 \begin{equation}
S = \begin{pmatrix}
1 & 0 & 0\\
0 & 0  & 1\\
0& 1 & 0\\\\
\end{pmatrix}
\end{equation}
A small explicit breaking of $ \mu-\tau  $ symmetry is put by hand, by inheriting the property,
\begin{equation}
h_{22} \neq h_{33}
\end{equation} 
This introduces CP violation in PMNS matrices and results in $ \theta_{13} $ being non zero. The amount of breaking used here for generating non zero CP asymmetry producing a measurable CP violating phase via Dirac neutrino Yukawa Couplings used from {\color{blue}\cite{kod}} is
\begin{equation}
\frac{h_{22}-h_{33}}{h_{22}+h_{33}} =  0.0045
\end{equation}
Exact $ \mu-\tau $ symmetry imposes $ Sin^{2}\theta_{23} = 0.$ One can break the symmetry spontaneously through the vev of the 120 plet and by use of Eq. (22). A required amount of CP violating phase $ \delta_{CP} $ is generated by explicitly breaking $ \mu-\tau $ symmetry. This assumption by using Eq.(22) leads to $ Sin^{2}\theta_{23} \sim 0.42-0.63$ and $ Sin^{2}\theta_{13} > 0.006$. All these remarks enables one to use Dirac neutrino Yukawa Couplings mass matrices reproduced by explicit use of broken $\mu-\tau$ symmetry embedded in Eq. (22). 
\section{Numerical Analysis}

In this section, numerical analysis has been carried out. Firstly, the free parameter called the lightest $ \nu $ mass, the $ \nu $ oscillation parameters like reactor angle $ \theta_{13} $  Dirac CPV phase, $ \delta_{CP} $, Majorana phases $ \alpha_{21} $, $ \alpha_{31} $ in broken $ \mu-\tau $ symmetry, type I Seesaw model are scanned to search for dependance of  $ \delta_{CP} $ phase on $ \theta_{13} $, $ m_{1} $ ($ m_{3} $), Jarkslog invariant $ J_{CP} $, effective mass for $ )\nu\beta\beta $ decay in case of normal ordering (inverted ordering) in the context of producing correct baryon asyymetry of the Universe {\color{blue}\cite{9}}. We use best fit values of $ \nu $ oscillation parameters. The two mass square differences $\Delta m^{2}_{12}$, $\Delta m^{2}_{13}$ are embedded in neutrino mixing matrix so we are left out with lightest $ \nu $ mass as only free parameter in this model. In the chargd lepton basis, we parameterize the PMNS matrix $U_{PMNS}$, by diagonalizing the neutrino mass matrix $m_{\nu}$ in terms of three mixing angles $\theta_{ij}$ ($i, j = 1, 2, 3$; $i < j$), one CP violating Dirac CPV phase $ \delta_{CP} $, and two Majorana phases ($ \alpha_{21} $ and $  \alpha_{31}$) as follows:
\begin{equation}
U^{\ast}P^{\ast}m_{\nu}P^{\dagger}U^{\dagger} = m_{\nu}^{D}
\end{equation}
The Pontecorvo-Maki-Nakagawa-Sakata (PMNS) matrix is UP, where U is 
\begin{equation}
U = \begin{pmatrix}
c_{12}c_{13} & s_{12}c_{13} & s_{13}e^{-i\delta}\\
-s_{12}c_{23}-c_{12}s_{23}s_{13}e^{i\delta} & c_{12}c_{23}-s_{12}s_{23}s_{13}e_{i\delta}& s_{23}c_{13}\\
s_{12}s_{23}-c_{12}c_{23}s_{13}e^{i\delta} & -c_{12}s_{23}-s_{12}c_{23}s_{13}e^{i\delta} & c_{23}c_{13}\\
\end{pmatrix}
\end{equation}
where, $\theta_{12} = 33.82^{0}, \theta_{23} = 48.3^{0}(48.6^{0}), \theta_{13} = 8.61^{0}(8.65^{0})$ \cite{pdg} in case of normal hierarchy (inverted hierarchy) are the solar, atmospheric and reactor angles respectively. 
The Majorana phases reside in P, where 
\begin{equation}
P = \text{diag} \begin{pmatrix} 1 & e^{i\alpha_{21}} & e^{i(\alpha_{31} + \delta)}\end{pmatrix}
\end{equation}
We have taken complex and orthogonal matrix $R = U_{PMNS}$, in terms of user defined Dirac neutrino Yukawa Couplings defined in Eq. (13) in order to produce correct baryon asymmetry of the Universe.
\par 
For the Normally ordered light $\nu$ masses, we have 
\begin{equation}
M_{R}^{diag} = \text{diag} (M_{1} , M_{2}, M_{3} )
= M_{1} \text{diag}(1, \frac{M_{2} }{M_{1}} , \frac{M_{3}}{M_{1}} )
= M_{1}\text{diag}(1, \frac{m_{1}}{m_{2}} ,\frac{m_{1}}{m_{3}})
\end{equation}
With $m_{1} \in [10^{-6} eV , 10^{-1} eV],
and, m_{2}^{2} - m_{1}^{2} = 7.39\times 10^{-5} eV^{2}, m_{3}^{2} - m_{1}^{2} = 2.48 \times 10^{-3} eV^{2} $
as is evident from the $\nu$ oscillation data \cite{pdg}, $m_{1}$ being the lightest of three  $ \nu $ masses. For the inverted ordered light $ \nu $ masses, we have 
\begin{equation}
M_{R}^{diag} = \text{diag} (M_{1} , M_{2}, M_{3} )
= M_{1} \text{diag}(1, \frac{M_{2} }{M_{1}} , \frac{M_{3}}{M_{1}} )
= M_{1}\text{diag}(1, \frac{m_{1}*m_{3}}{m_{2}^{2}} ,\frac{m_{1}}{m_{2}})
\end{equation}
with $ m_{3} $ being the lightest of three  $ \nu $ masses. Here we take $M_{1} \sim 10^{12}$ GeV. For normal ordering, the choices of lightest neutrino mass is $m_{1} = 0.07118 eV$ whereas for inverted ordering, the choice of lightest neutrino mass is $ m_{3} = 0.0657 eV$. This sustainable allowance of $ m_{1} (m_{3}) = 0.07(0.065)
eV$ signifies a neutrino mass spectrum where the sum of absolute neutrino masses
lies below the cosmological upper bound, $ \sum_{i} m (\nu_{i}) < 0.23 \text{eV}$ \cite{An}. 
Next random scan of the $ \nu $ mixing matrix parameter space for NH, IH in order to produce correct baryon asymmetry of the Universe $5.8\times 10^{-10} < Y_{B} < 6.6 \times 10^{-10} $ is performed in the following 3 $ \sigma $ranges of $ \delta_{CP} $ with respect to the tabulated $ \chi^{2} $ map of the SuperKamiokande analysis of the data within $ \Delta \chi^{2} = 6.2$ {\color{blue}\cite{pdg}}:
$$m_{1}(m_{3} ) \in [10^{-6}\hspace{.1cm} eV, 0.1 \hspace{.1cm} eV]\hspace{.1cm} ([10^{-6} \hspace{.1cm} eV, 0.1 \hspace{.1cm} eV] )$$
$$\delta_{CP} \in [ 141, 370]\hspace{0.2cm} \text{for Normal ordering}$$
$$\delta_{CP} \in [ 205, 354] \hspace{0.2cm}\text{for inverted ordering}$$
$$\theta_{13} \in [ 7.9, 8.9]\hspace{0.2cm} \text{for Normal ordering for tabulated $\Delta\chi^{2} = 9.5$}$$
$$\theta_{13} \in [ 8.0, 9.0] \hspace{0.2cm}\text{for inverted ordering for tabulated $\Delta\chi^{2} = 9.5$}$$
$$\theta_{23} \in [ 40.8, 51.3] \hspace{0.2cm}\text{for normal ordering for tabulated $\Delta\chi^{2} = 6.2$}$$
While doing parameter scan, we find  favoured values of lightest $ \nu $ mass, dirac CPV phase $ \delta_{CP} $, for producing correct baryon asymmetry of the Universe, $5.8\times 10^{-10} < \eta < 6.6 \times 10^{-10} $. 
\par 
The lepton flavor effects are significant if the lightest right handed Majorana neutrino mass $ M_{\nu_{R1}} $ is below $ 10^{12}$ GeV. Here $ M_{1} = 10^{12}$ GeV. In the type I seesaw mechanism one can always find the right handed neutrino mass matrix as;
\begin{equation}
M_{\alpha \beta} = -(m_{D}^{T})_{\alpha I}(M_{IJ}^{-1})(m_{D})_{J \beta}
\end{equation}
where $(m_{D})_{\alpha I}$ is the Dirac mass matrix. We consider a Dirac neutrino mass matrix defined in Eq. (13). Here when we fix $ m_{1} (m_{3}), Y_{\nu}, M_{R}$
 the remaining free parameter in the neutrino sector within our broken $ \mu-\tau $ framework is the leptonic CPV phase $ \delta_{CP} $. When we vary the CPV phase $ \delta_{CP} $ we compute the favoured regions of $m_{1}(m_{3})$. The
variations of leptonic CPV phase $ \delta_{CP} $ with $m_{1}(m_{3})$, $ \theta_{13} $, $ J_{CP} $, $ m_{ee} $  for $ 0\nu\beta\beta $  are shown in figures below.
\par
For global fit values of $ \nu $ oscillation parameters, we compute the Jarlskog invariant, $ \delta_{CP} $ given by PMNS matrix elements $ U_{\alpha i} $. We also compute the Jarkslog invariant for allowed values of $ \delta_{CP} $ phase, $ \theta_{13} $ and lightest $ \nu $ mass explored here in this work for both normal ordering and inverted ordering. 
\begin{equation}
J_{CP} = Im[U_{e1}U_{\mu2}U^{*}_{e2}U^{*}_{\mu1}] = s_{23}c_{23}s_{12}c_{12}s_{13}c^{2}_{13}sin\delta_{CP}
\end{equation}
We also calculate the favourable space of the effective mass for $ 0\nu\beta\beta $ decay 
for favourable values of $ \delta_{CP} $ phase, $ \theta_{13} $ and lightest $ \nu $ mass  given by, 
\begin{equation}
m_{ee} = |m_{1}c^{2}_{12}c^{2}_{13} + m_{2}s^{2}_{12}c^{2}_{13}e^{i\alpha_{21}} + m_{3}s^{2}_{13}e^{i({\alpha_{31}-2\delta_{CP}})}|
\end{equation}

\begin{center}
\begin{figure*}[htbp]
\centering{
\begin{subfigure}[]{\includegraphics[height=7.4cm,width=7.9cm]{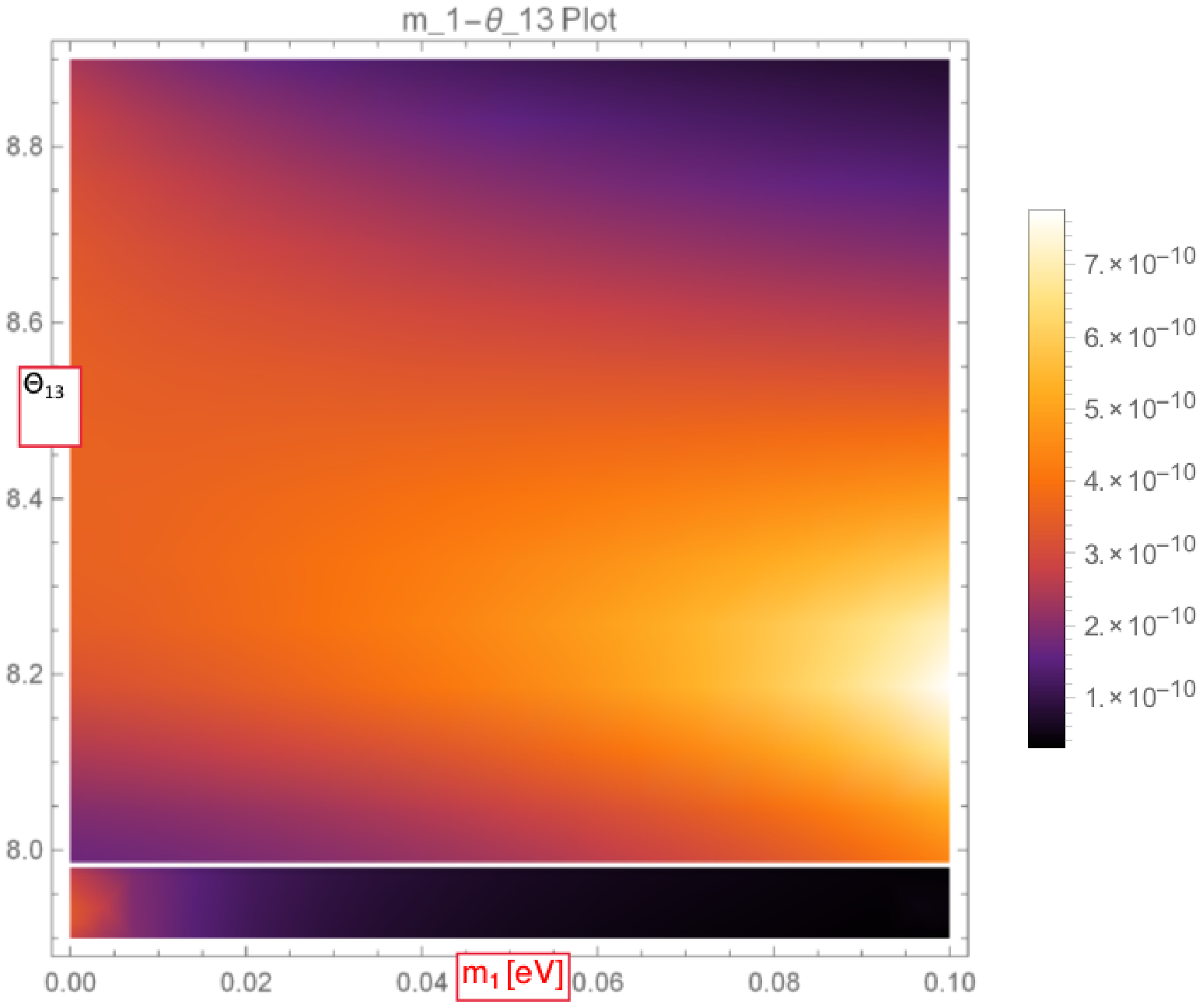}}\end{subfigure}
\begin{subfigure}[]{\includegraphics[height=	7.4cm,width=7.9cm]{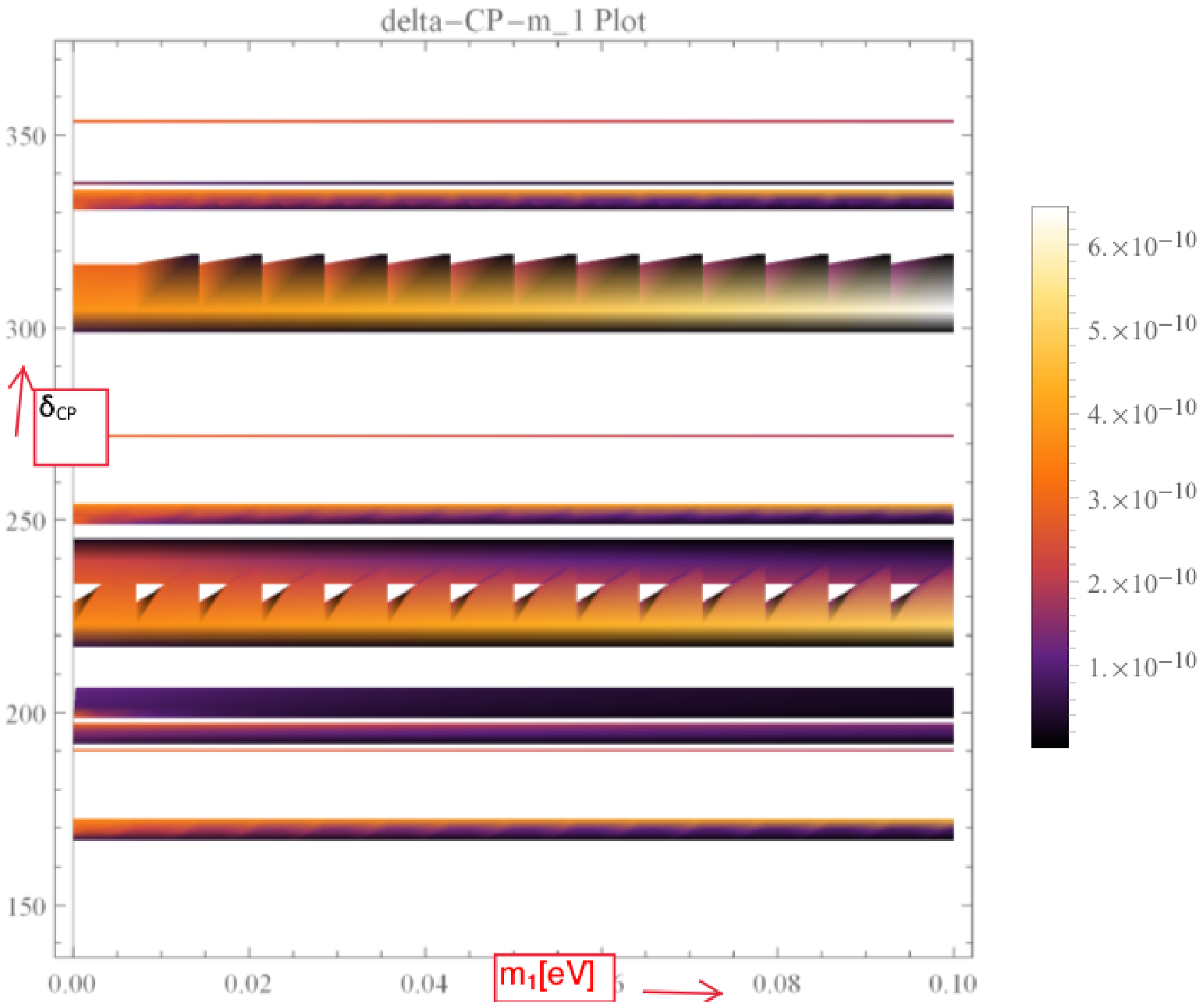}}\end{subfigure}\\

\caption{Predictions in broken $ \mu-\tau $ symmetry model for Normal ordering. The left panel: predicted favoured values of $({m_{1}, \theta_{13}})$ plane for best fit values of $ \delta_{CP} = 222^{0}$ with $\Delta \chi^{2} =6.2$ w/o SK-ATM {\color{blue}\cite{pdg}} ( allowed by updated values of correct baryon asymmetry of the Universe) as a result of contribution of type I Seesaw mechanism to neutrino mass matrix. Similarly, The right panel: predicted favoured values $({m_{1}, \delta_{CP}})$ plane, for best fit value of $ \theta_{13} = 8.41 $ with $\Delta \chi^{2}=9.5${\color{blue}\cite{pdg}}}}. 
\label{fig:1}
\end{figure*}
\end{center}

\begin{center}
\begin{figure*}[htbp]
\includegraphics[height=8cm,width=11cm]{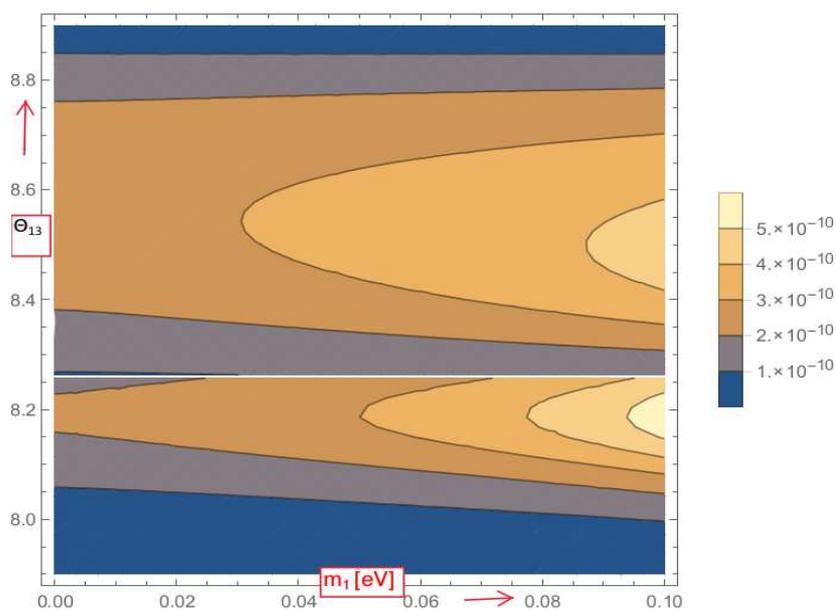}

\caption{Predictions in broken $ \mu-\tau $ symmetry model for Normal ordering: Contour plot for predicted favoured values of $({m_{1}, \theta_{13}})$ plane for best fit values of $ \delta_{CP} = 222^{0}$ of $\Delta_{\chi^{2}}=6.2$ w/o SK-ATM {\color{blue}\cite{pdg}} ( allowed by updated values of correct baryon asymmetry of the Universe) as a result of contribution of type I Seesaw mechanism to neutrino mass matrix.}
\label{fig:1}
\end{figure*}
\end{center}

\begin{center}
\begin{figure*}[htbp]
\centering{
\begin{subfigure}[]{\includegraphics[height=6.9cm,width=7.9cm]{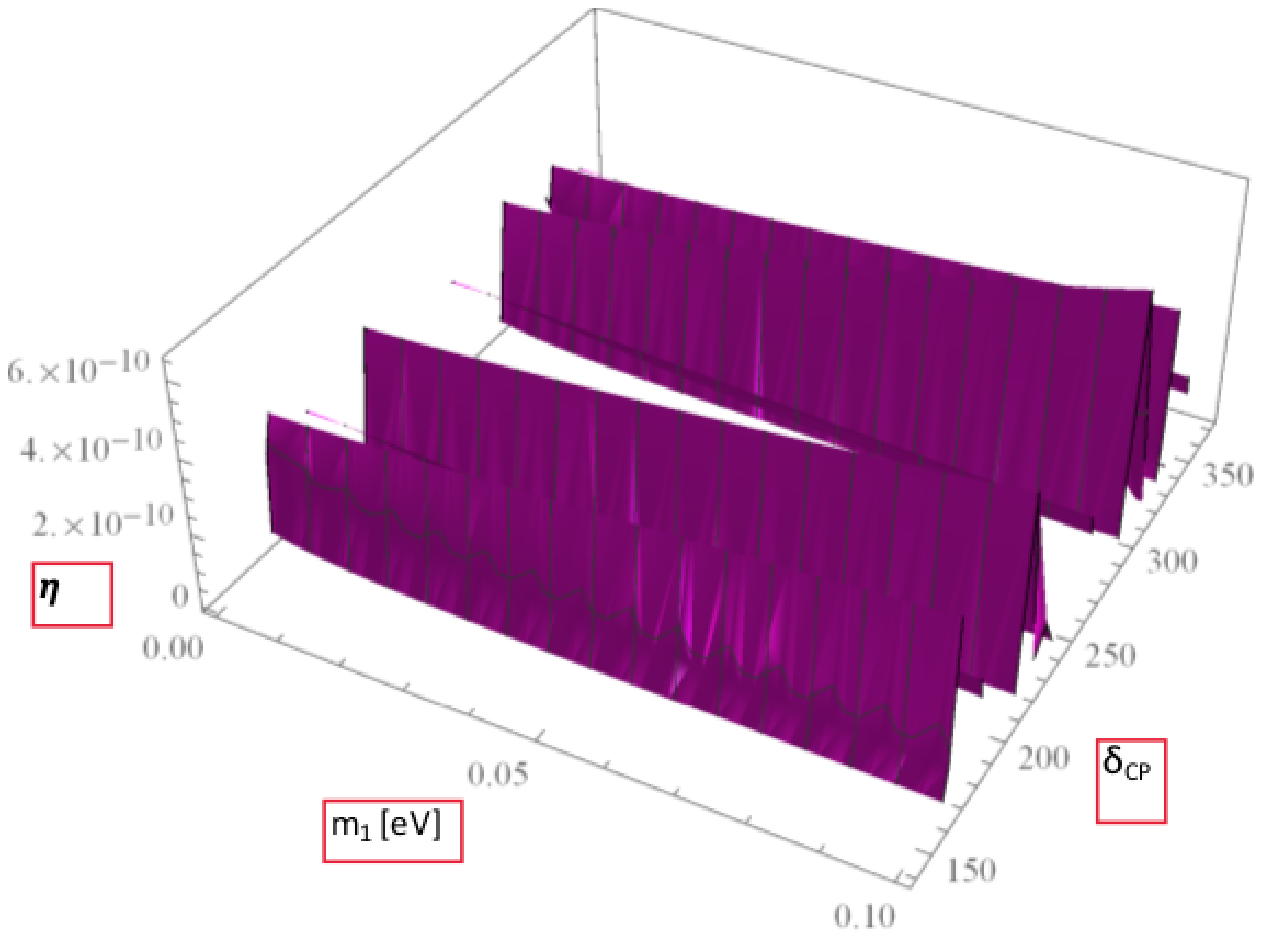}}\end{subfigure}
\begin{subfigure}[]{\includegraphics[height=	6.9cm,width=7.9cm]{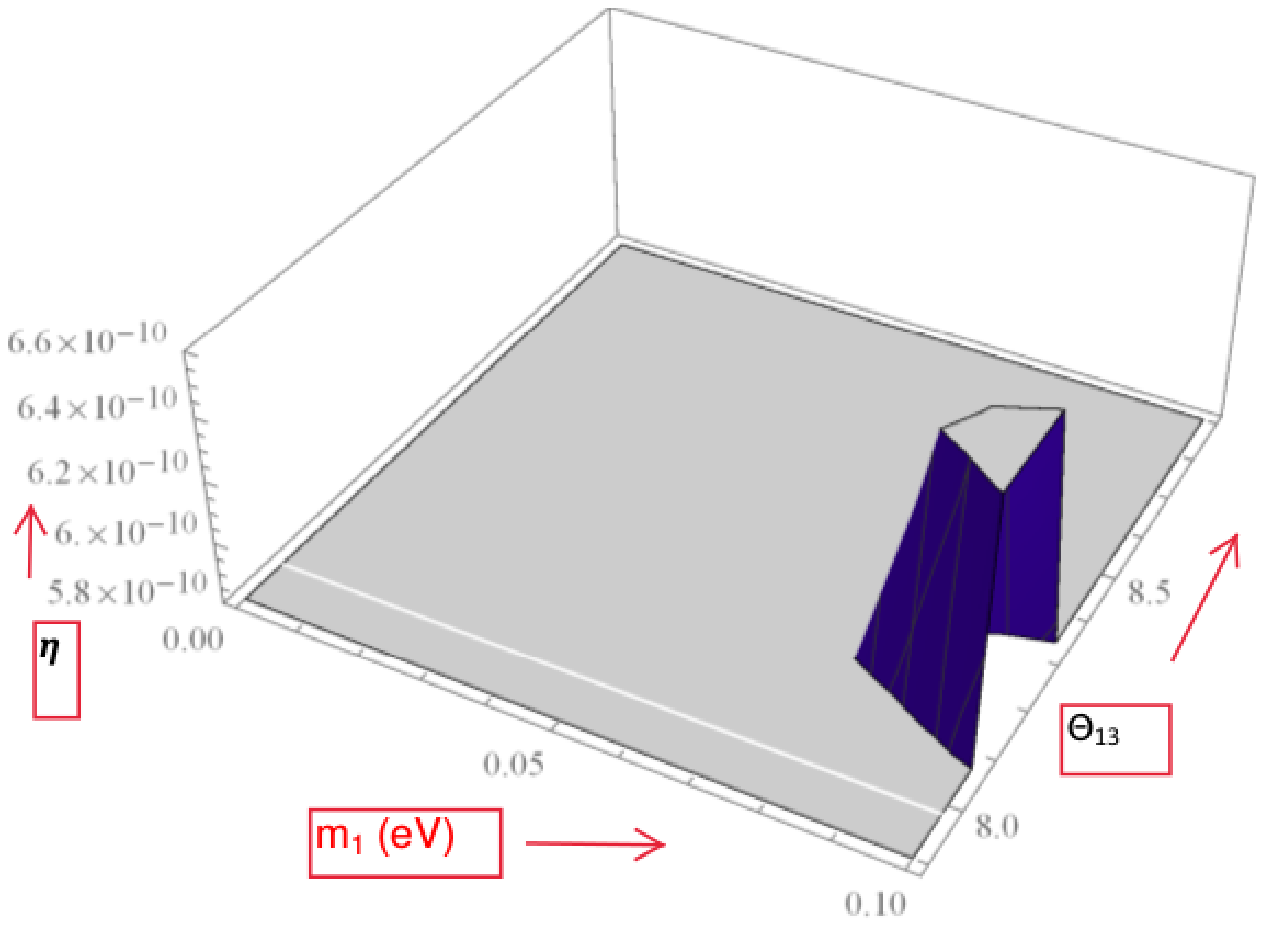}}\end{subfigure}\\

\caption{Predictions in broken $ \mu-\tau $ symmetry model for Normal ordering: The left panel: three dimensional plot of  preferred values of $({m_{1}, \delta_{CP}})$ plane for best fit values of $ \theta_{13} = 8.41 $ of $\Delta \chi^{2}=9.5 $ w/o SK-ATM {\color{blue}\cite{pdg}} (in the light of recent ratio of the baryon to photon density bounds $5.8\times 10^{-10} < \eta < 6.6 \times 10^{-10} $) as a result of contribution of type I Seesaw mechanism to neutrino mass matrix. Similarly, The right panel:  three dimensional plot of  favourable values of  $({m_{1}, \theta_{13}})$ plane, for best fit value of $ \delta_{CP} = 222^{0}$ of $\Delta \chi^{2}=6.2$  {\color{blue}\cite{pdg}} (in the light of recent ratio of the baryon to photon density bounds $5.8\times 10^{-10} < \eta < 6.6 \times 10^{-10} $).}}
\label{fig:1}
\end{figure*}
\end{center}

\begin{center}
\begin{figure*}[htbp]
\centering{
\begin{subfigure}[]{\includegraphics[height=6.9cm,width=7.9cm]{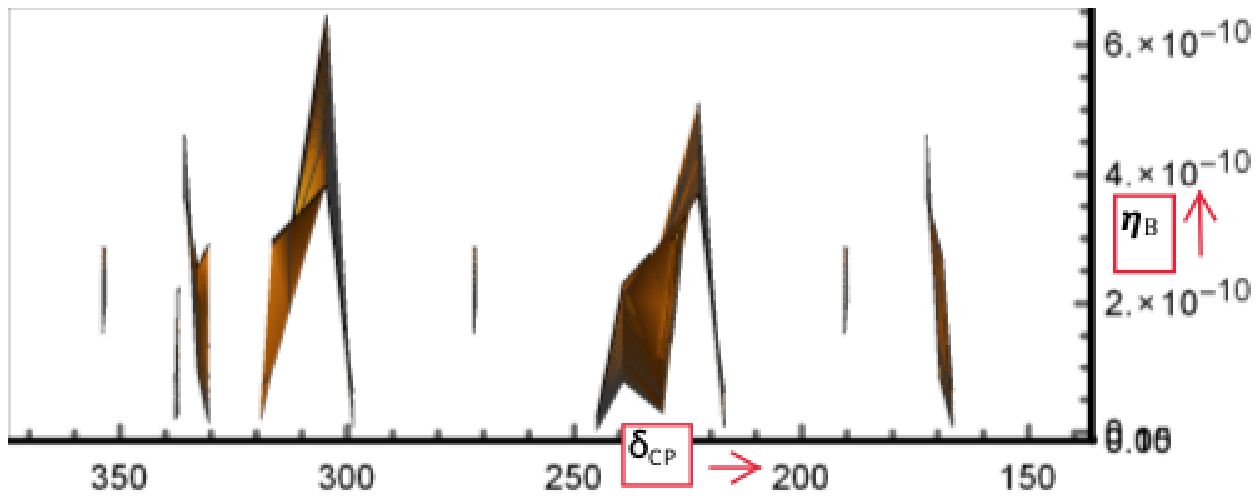}}\end{subfigure}
\begin{subfigure}[]{\includegraphics[height=	6.9cm,width=7.9cm]{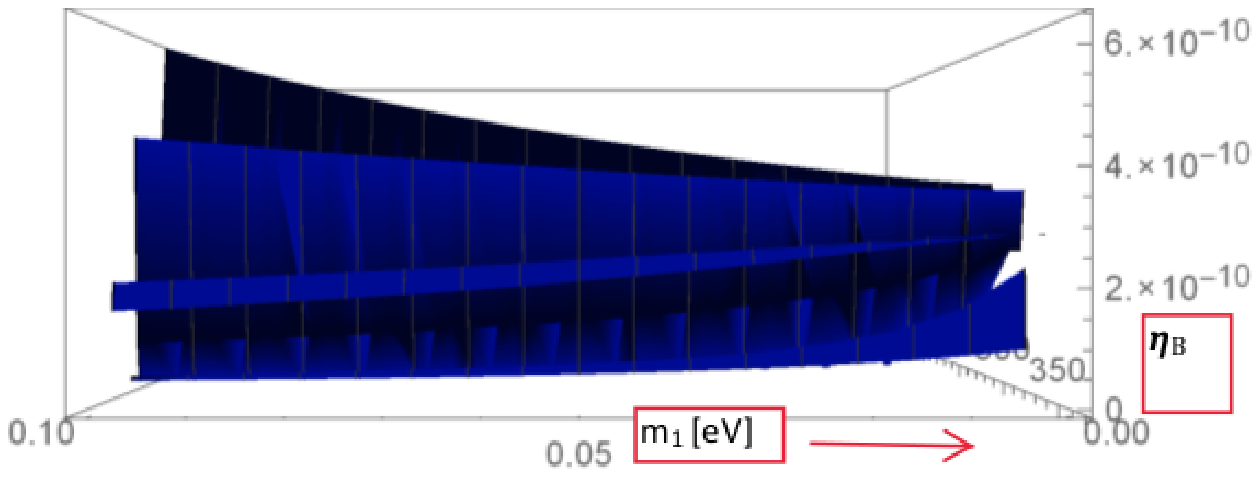}}\end{subfigure}\\

\caption{ Predictions in broken $ \mu-\tau $ symmetry model for Normal ordering: The left panel: preferred values of $\delta_{CP}$ for best fit values of $ \theta_{13} = 8.41 $ of $\Delta \chi^{2} = 9.5 $ w/o SK-ATM {\color{blue}\cite{pdg}} (in the light of recent ratio of the baryon to photon density bounds $ 5.8\ times 10^{-10} < \eta < 6.6 \times 10^{-10} $) as a result of contribution of type I Seesaw mechanism to neutrino mass matrix. Similarly, the right panel: favoured values of lightest neutrino mass, $m_{1}$, for best fit value of $ \delta_{CP} = 222^{0}$ of $\Delta \chi^{2} = 6.2$  {\color{blue}\cite{pdg}} (in the light of recent ratio of the baryon to photon density bounds $ 5.8 \times 10^{-10} < \eta < 6.6 \times 10^{-10} $).}}
\label{fig:1}
\end{figure*}
\end{center}

\begin{center}
\begin{figure*}[htbp]
\centering{
\begin{subfigure}[]{\includegraphics[height=5.9cm,width=7.9cm]{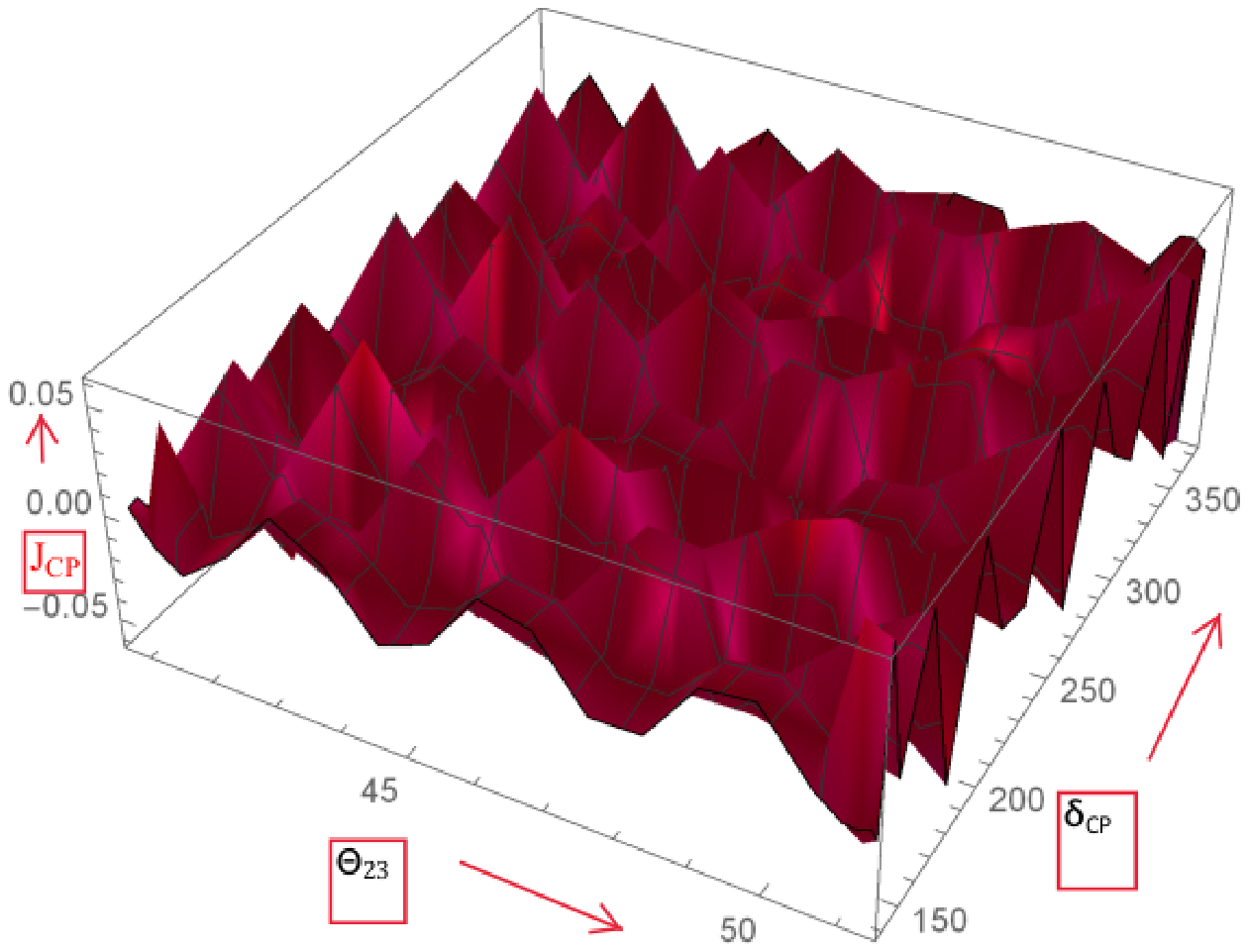}}\end{subfigure}
\begin{subfigure}[]{\includegraphics[height=	5.9 cm,width=9.9 cm]{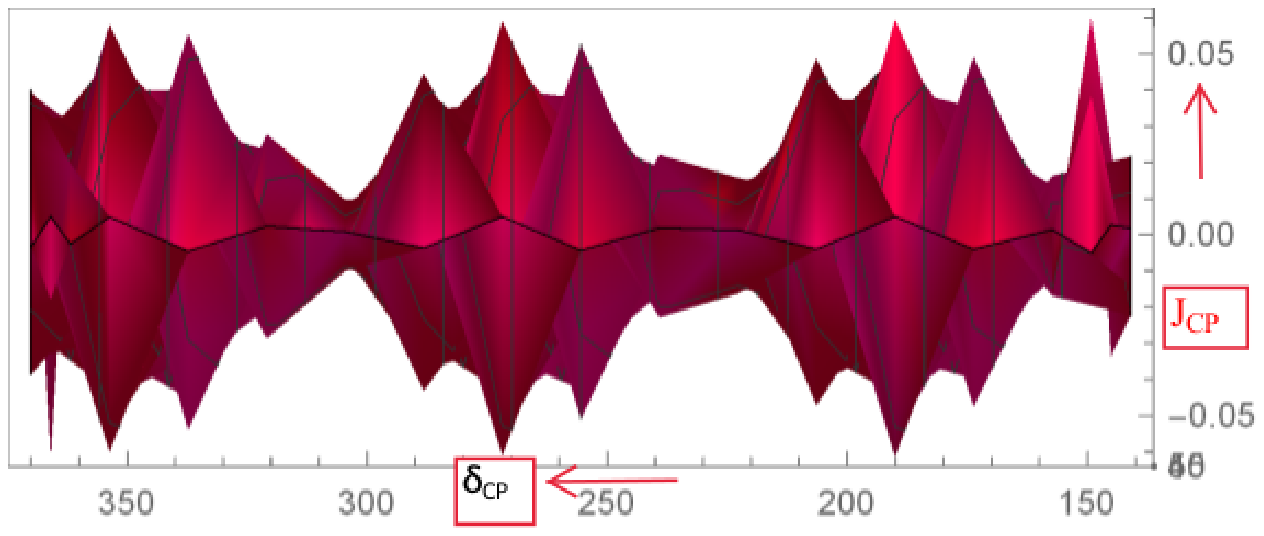}}\end{subfigure}\\

\caption{Predictions in broken $ \mu-\tau $ symmetry model for Normal ordering: The right panel: preferred three dimensional regions of ($ \delta_{CP} $, $ \theta_{23} $, $ J_{CP} $) plane for best fit values of $ \theta_{13} = 8.41 $ of $\Delta \chi^{2} = 9.5 $ w/o SK-ATM {\color{blue}\cite{pdg}}. The left panel: allowed two dimensional space of ($ \delta_{CP} $, $ J_{CP} $) plane for best fit values of $ \theta_{13} = 8.41 $ of $\Delta \chi^{2} = 9.5 $ w/o SK-ATM {\color{blue}\cite{pdg}}.}}
\label{fig:1}
\end{figure*}
\end{center}

\begin{center}
\begin{figure*}[htbp]
\includegraphics[height=8cm,width=11cm]{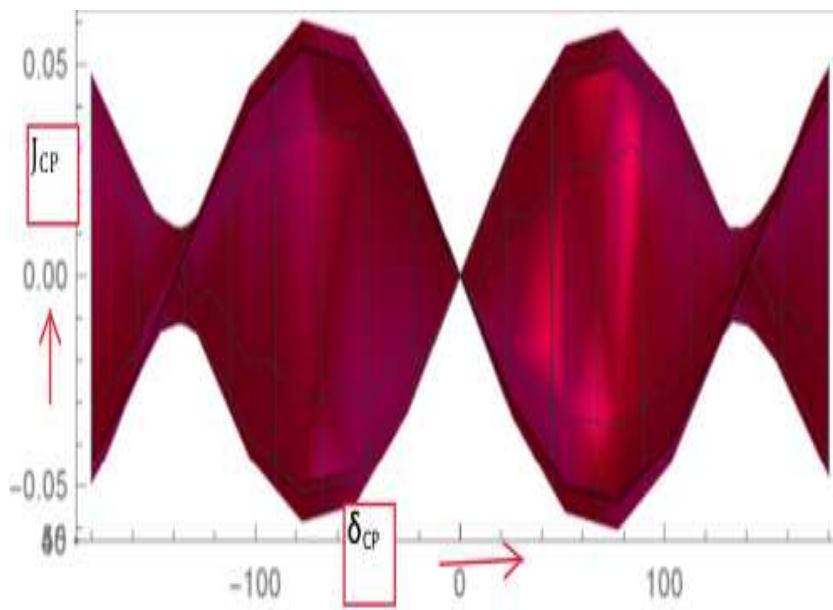}

\caption{Predictions in broken $ \mu-\tau $ symmetry model for Normal ordering: The right panel: allowed two dimensional space of ($ \delta_{CP} $, $ J_{CP} $ ) plane for absolute range of $ \delta_{CP} \in [-180,+180] $.}
\label{fig:1}
\end{figure*}
\end{center}

\begin{center}
\begin{figure*}[htbp]
\centering{
\begin{subfigure}[]{\includegraphics[height=5.9cm,width=7.9cm]{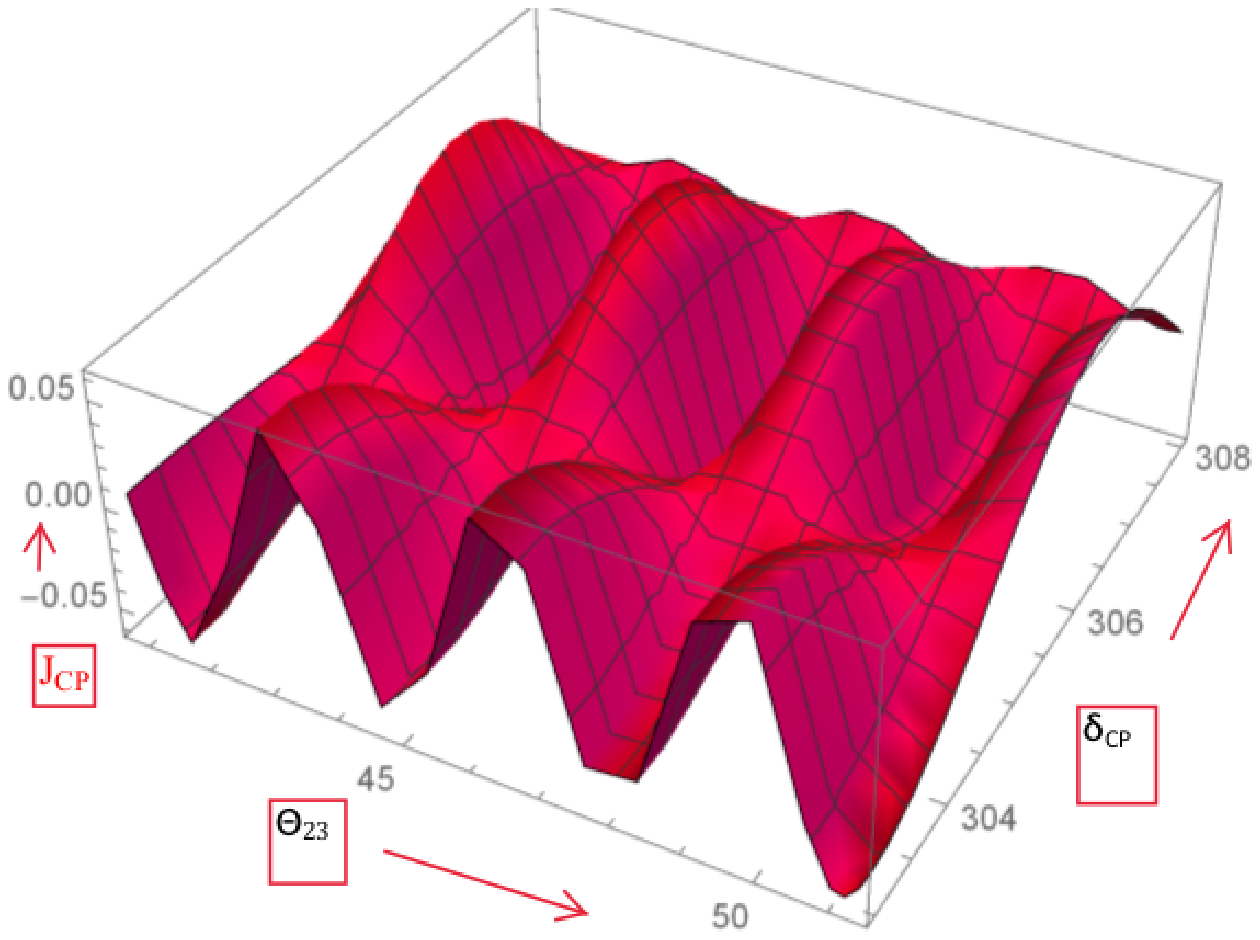}}\end{subfigure}
\begin{subfigure}[]{\includegraphics[height=	5.9 cm,width=7.9 cm]{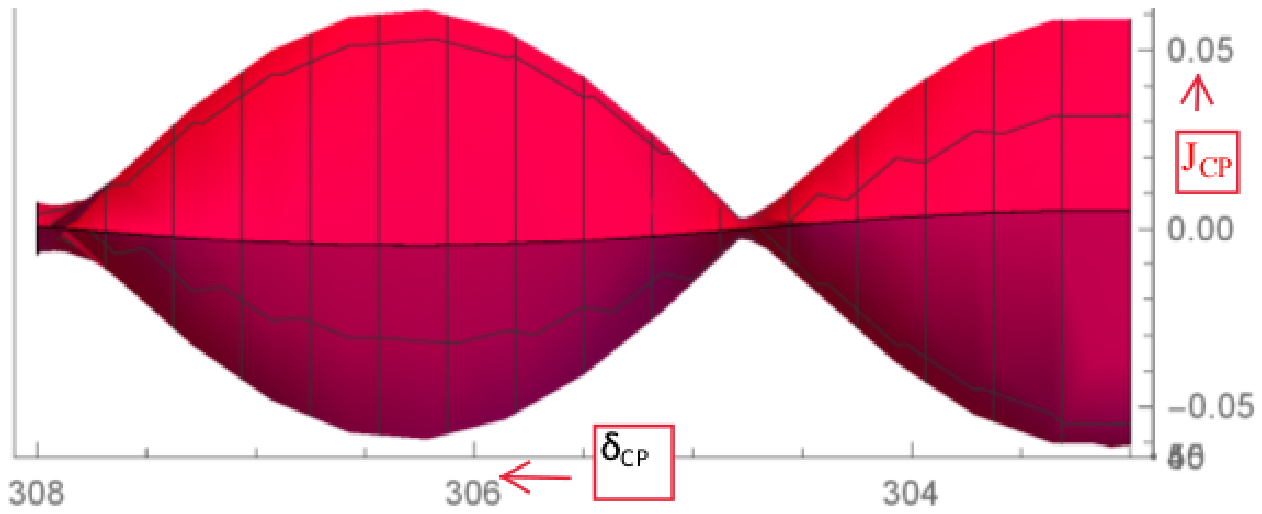}}\end{subfigure}\\

\caption{Predictions in broken $ \mu-\tau $ symmetry model for Normal ordering: The right panel: preferred three dimensional regions of ($ \delta_{CP} $, $ \theta_{23} $, $ J_{CP} $) plane for favoured values of $ \delta_{CP} \in [303, 308]$ (in the light of recent ratio of the baryon to photon density bounds $ 5.8 \times 10^{-10} < \eta < 6.6 \times 10^{-10} $) for best fit values of $ \theta_{13} = 8.41 $ of $\Delta \chi^{2} = 9.5 $ w/o SK-ATM {\color{blue}\cite{pdg}}. The left panel: allowed two dimensional space of ($ \delta_{CP} $, $ J_{CP} $) plane for favoured values of $ \delta_{CP} \in [303, 308]$ (in the light of recent ratio of the baryon to photon density bounds $ 5.8 \times 10^{-10} < \eta < 6.6 \times 10^{-10} $) for best fit values of $ \theta_{13} = 8.41 $ of $\Delta \chi^{2} = 9.5 $ w/o SK-ATM {\color{blue}\cite{pdg}} (in the light of recent ratio of the baryon to photon density bounds $ 5.8 \times 10^{-10} < \eta < 6.6 \times 10^{-10} $) .}}
\label{fig:1}
\end{figure*}
\end{center} 

\begin{center}
\begin{figure*}[htbp]
\centering{
\begin{subfigure}[]{\includegraphics[height=5.9cm,width=7.9cm]{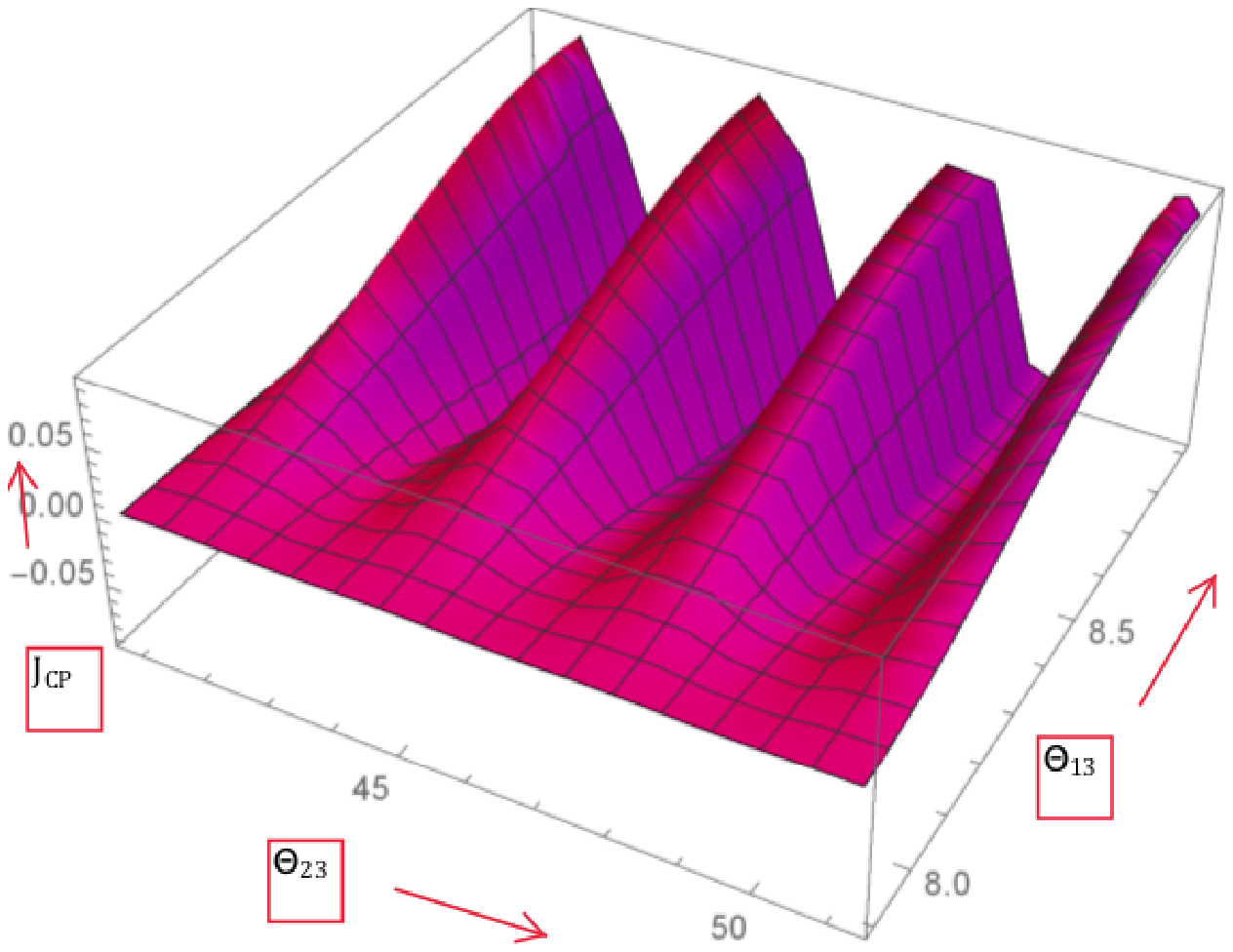}}\end{subfigure}
\begin{subfigure}[]{\includegraphics[height=	5.9 cm,width=7.9 cm]{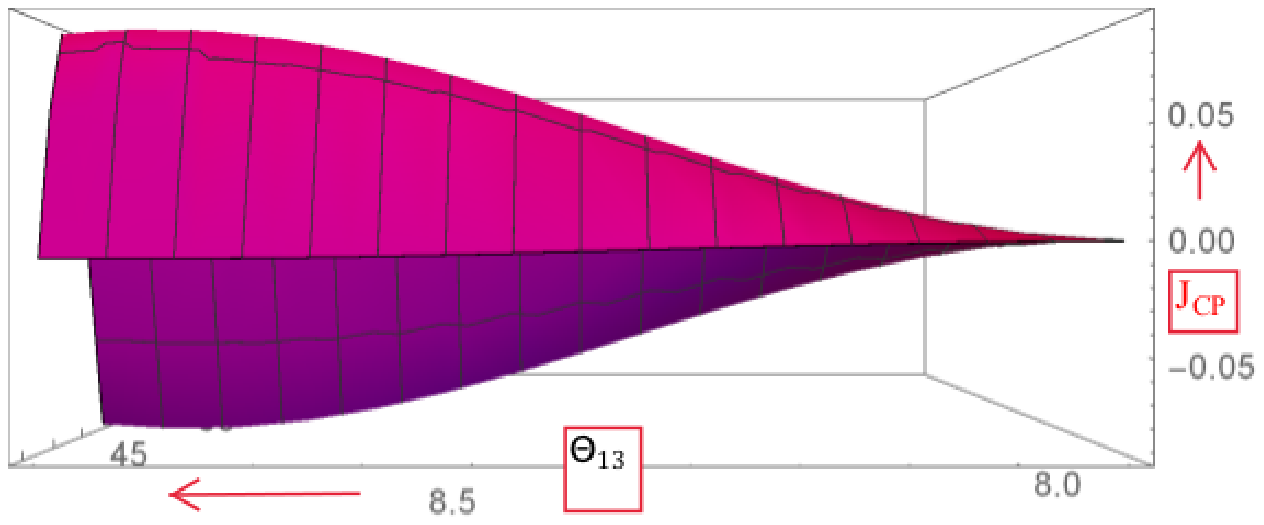}}\end{subfigure}\\

\caption{Predictions in broken $ \mu-\tau $ symmetry model for Normal ordering: The right panel: preferred three dimensional regions of ($ \theta_{13} $, $ \theta_{23} $, $ J_{CP} $) plane for best fit values of $ \delta_{CP} = 222^{0} $   of $\Delta \chi^{2} = 6.2 $ w/o SK-ATM {\color{blue}\cite{pdg}}. The left panel: allowed two dimensional space of ($ \theta_{13} $, $ J_{CP} $) plane for best fit values of $ \delta_{CP} = 222^{0} $ of $\Delta \chi^{2} = 6.2 $ w/o SK-ATM {\color{blue}\cite{pdg}}}}
\label{fig:1}
\end{figure*}
\end{center}

\begin{center}
\begin{figure*}[htbp]
\includegraphics[height=7cm,width=10cm]{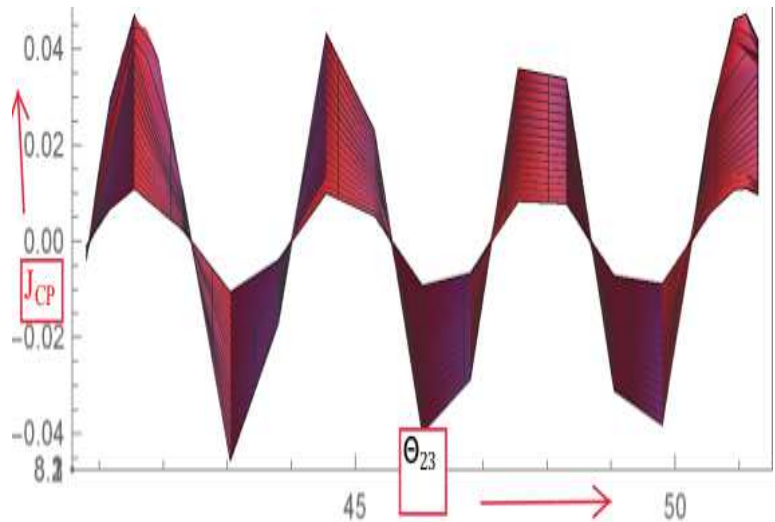}

\caption{Predictions in broken $ \mu-\tau $ symmetry model for Normal ordering: Variation of $ J_{CP} $ as a function of $ \theta_{23} $, $\theta_{23} \in [ 40.8, 51.3]$ {for normal ordering for tabulated $\Delta\chi^{2} = 6.2.$ \cite{pdg}.} }
\label{fig:1}
\end{figure*}
\end{center}

\begin{center}
\begin{figure*}[htbp]
\centering{
\begin{subfigure}[]{\includegraphics[height=6.9cm,width=7.9cm]{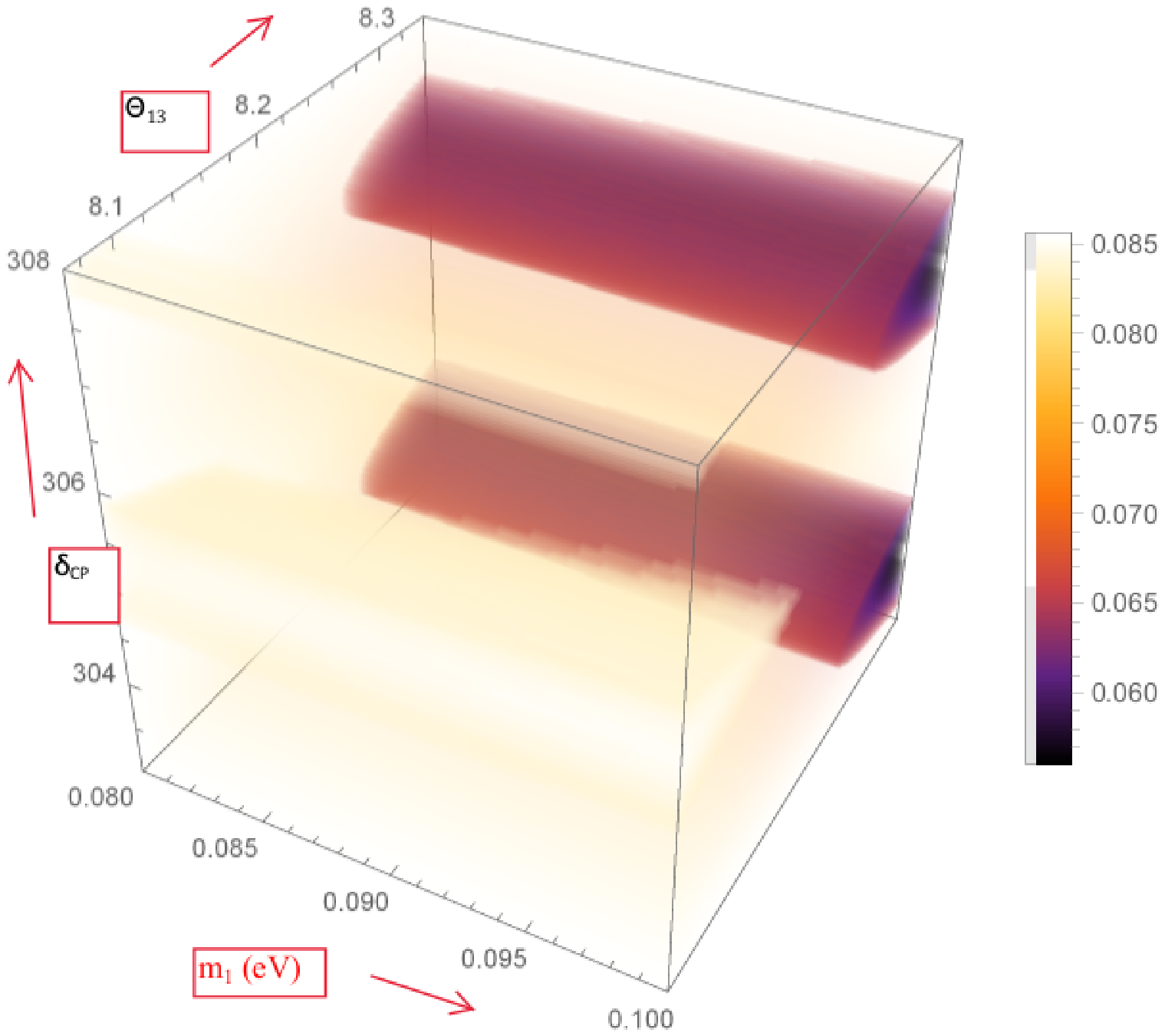}}\end{subfigure}
\begin{subfigure}[]{\includegraphics[height=	6.9 cm,width=7.9 cm]{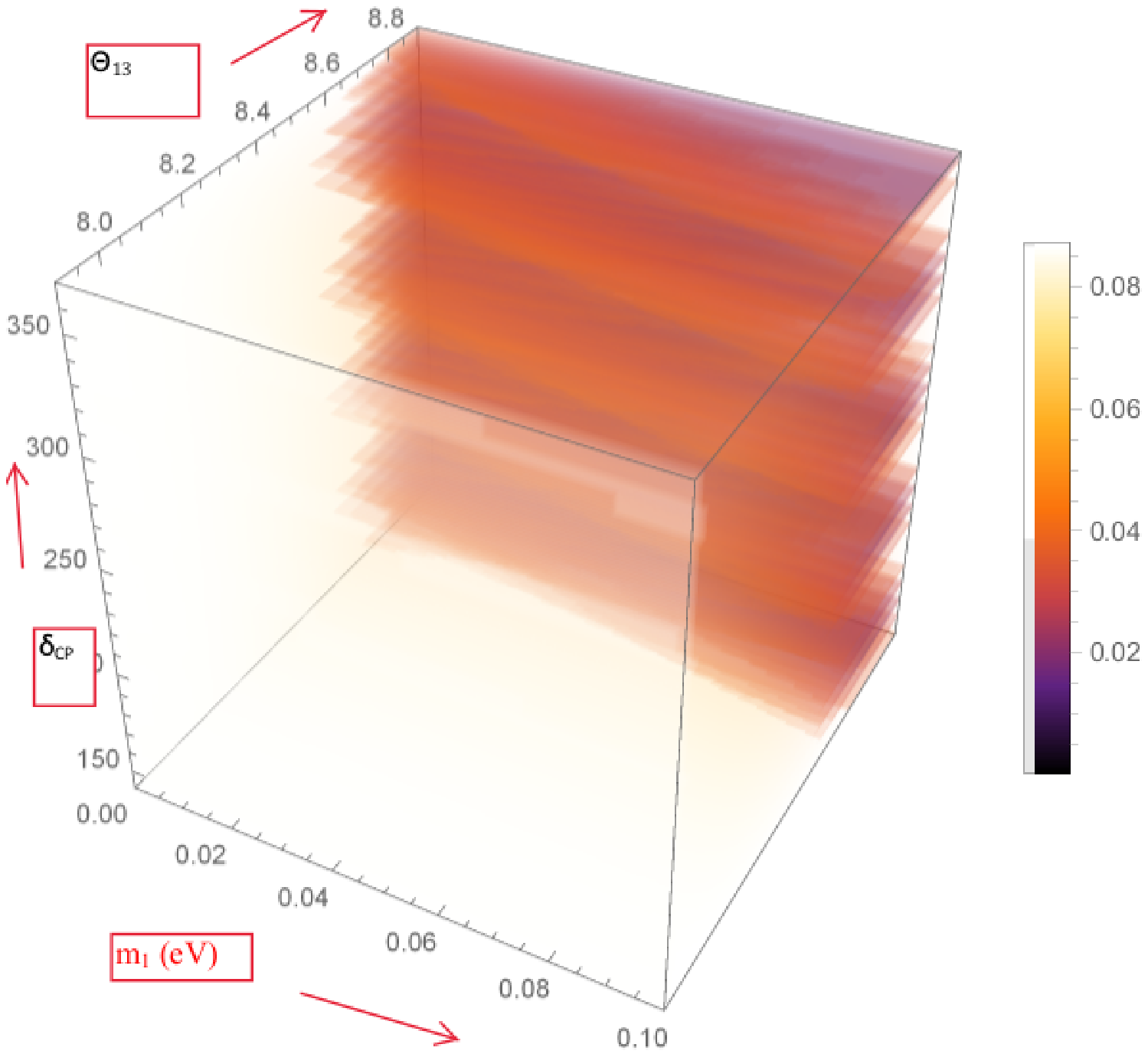}}\end{subfigure}\\

\caption{Predictions in broken $ \mu-\tau $ symmetry model for Normal ordering: The left panel depicts predicted three dimensional space of $(m_{1},\delta_{CP}, \theta_{13})$ for $ m_{ee} $ [eV], $ 0\nu\beta\beta $ decay for favoured values of $m_{1}$,$\delta_{CP}, \theta_{13}$ (in the light of recent ratio of the baryon to photon density bounds, $ 5.8 \times 10^{-10} < \eta < 6.6 \times 10^{-10} $). The right panel depicts predicted three dimensional space of ($m_{1},\delta_{CP}, \theta_{13}$) for $ m_{ee} $ [eV], $ 0\nu\beta\beta $ decay for values of $m_{1}$,$\delta_{CP}, \theta_{13}$ in the given three $ \sigma $ range, corresponding to $\Delta \chi^{2} = 6.2 $ and $\Delta \chi^{2} =9.5 $  w/o SK-ATM {\color{blue}\cite{pdg}}.}}
\label{fig:1}
\end{figure*}
\end{center}

\begin{center}
\begin{figure*}[htbp]
\centering{
\begin{subfigure}[]{\includegraphics[height=5.9cm,width=7.9cm]{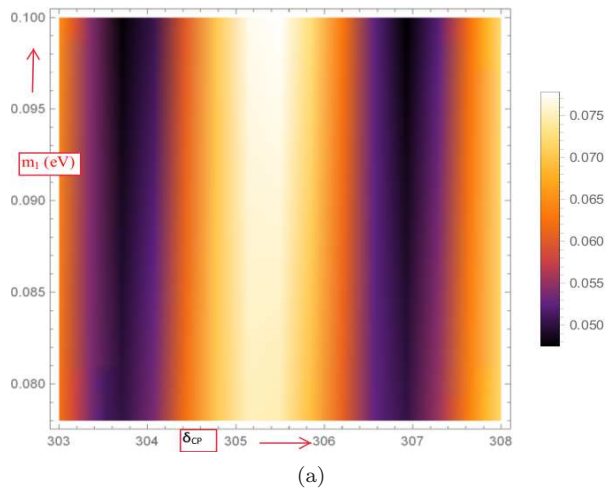}}\end{subfigure}\\

\caption{Predictions in broken $ \mu-\tau $ symmetry model for Normal ordering: depicts density plot of $(m_{1},\delta_{CP}$) for $ m_{ee}) $ [eV], $ 0\nu\beta\beta $ decay for favoured values of $m_{1}$,$\delta_{CP},\theta_{13}$ (in the light of recent ratio of the baryon to photon density bounds, $ 5.8 \times 10^{-10} < \eta < 6.6 \times 10^{-10} $). .}}
\label{fig:1}
\end{figure*}
\end{center}

\begin{center}
\begin{figure*}[htbp]
\centering{

\begin{subfigure}[]{\includegraphics[height=	5.9 cm,width=7.9 cm]{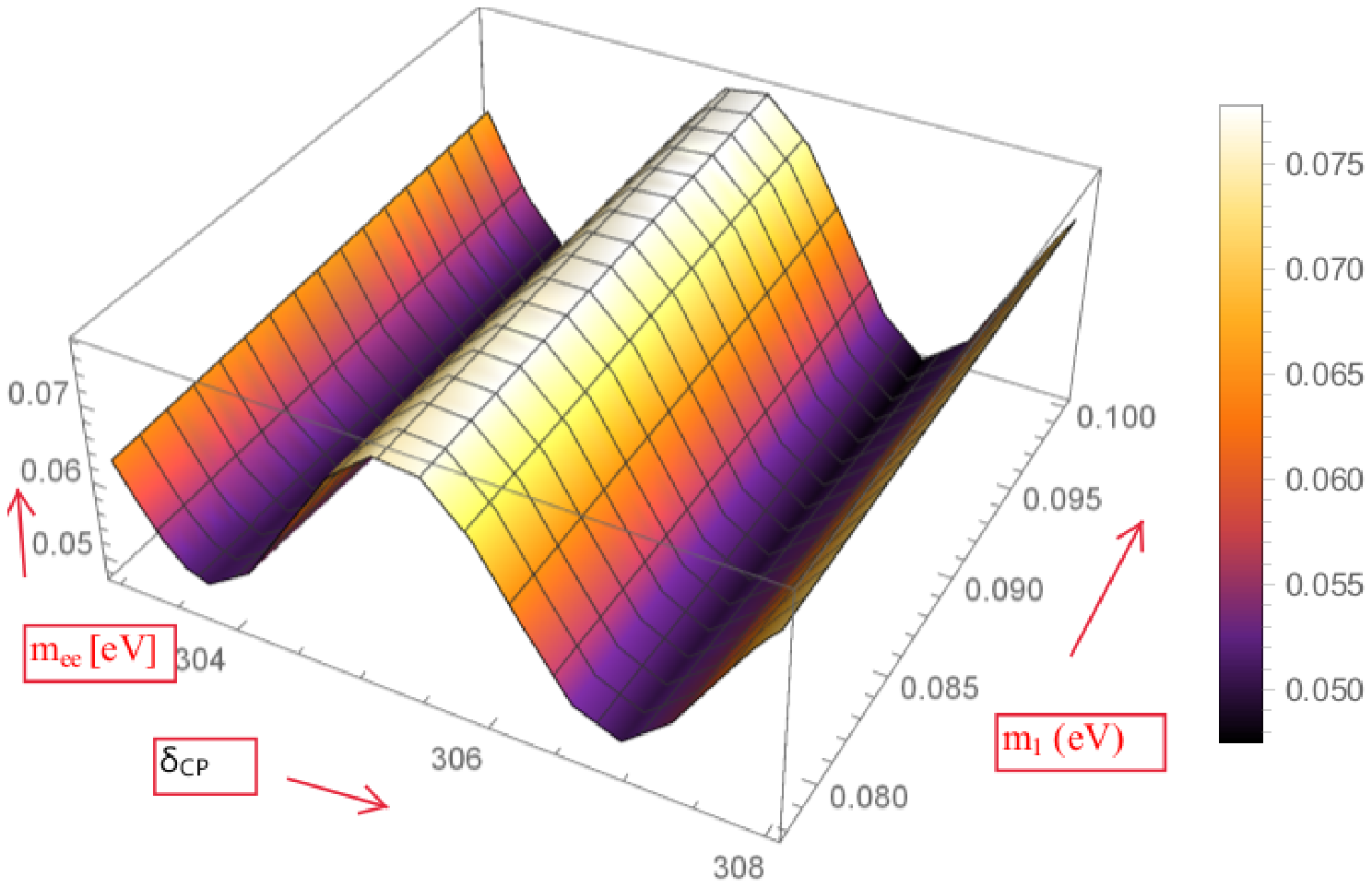}}\end{subfigure}
\begin{subfigure}[]{\includegraphics[height=	5.9 cm,width=7.9 cm]{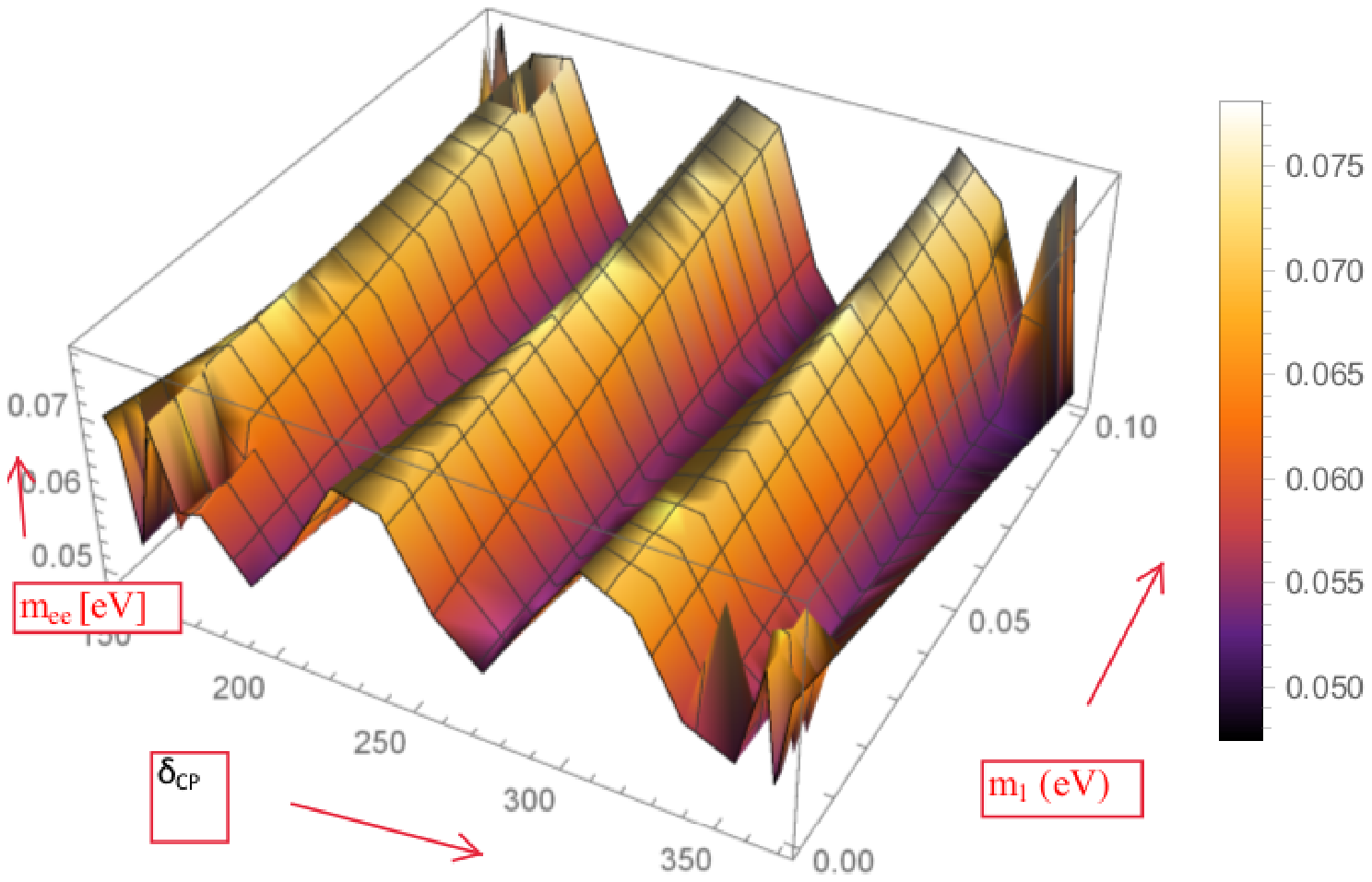}}\end{subfigure}\\
\caption{Predictions in broken $ \mu-\tau $ symmetry model for Normal ordering. Left panel depicts predicted three dimensional space of $(m_{ee},\delta_{CP}, m_{1})$ for $ m_{ee} $ [eV], $ 0\nu\beta\beta $ decay for favoured values of $m_{1}$,$\delta_{CP}, \theta_{13}$ (in the light of recent ratio of the baryon to photon density bounds, $ 5.8 \times 10^{-10} < \eta < 6.6 \times 10^{-10} $). Left panel depicts predicted three dimensional space of $(m_{ee},\delta_{CP}, m_{1})$ for $ m_{ee} $ [eV], $ 0\nu\beta\beta $ decay for favoured values of $m_{1}$,$\delta_{CP}, m_{ee}$ for values of $\delta_{CP}$ in the given three $ \sigma $ range, corresponding to $\Delta \chi^{2} =9.5 $ w/o SK-ATM {\color{blue}\cite{pdg}}}}
\label{fig:1}
\end{figure*}
\end{center}

\begin{center}
\begin{figure*}[htbp]
\includegraphics[height=7cm,width=10cm]{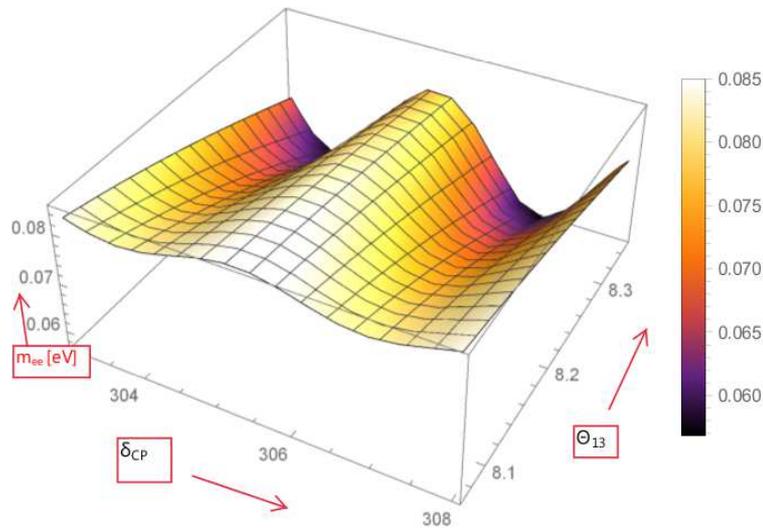}

\caption{The predicted three dimensional space of $(m_{ee},\delta_{CP}, \theta_{13})$ for $ m_{ee} $ [eV], $ 0\nu\beta\beta $ decay for favoured values of $m_{1}$,$\delta_{CP}, \theta_{13}$ (in the light of recent ratio of the baryon to photon density bounds, $ 5.8 \times 10^{-10} < \eta < 6.6 \times 10^{-10} $) for lightest $ \nu $ mass $ m_{1} = 0.07118 $ eV. }
\label{fig:1}
\end{figure*}
\end{center}

\begin{center}
\begin{figure*}[htbp]
\centering{
\begin{subfigure}[]{\includegraphics[height=5.4cm,width=6.9cm]{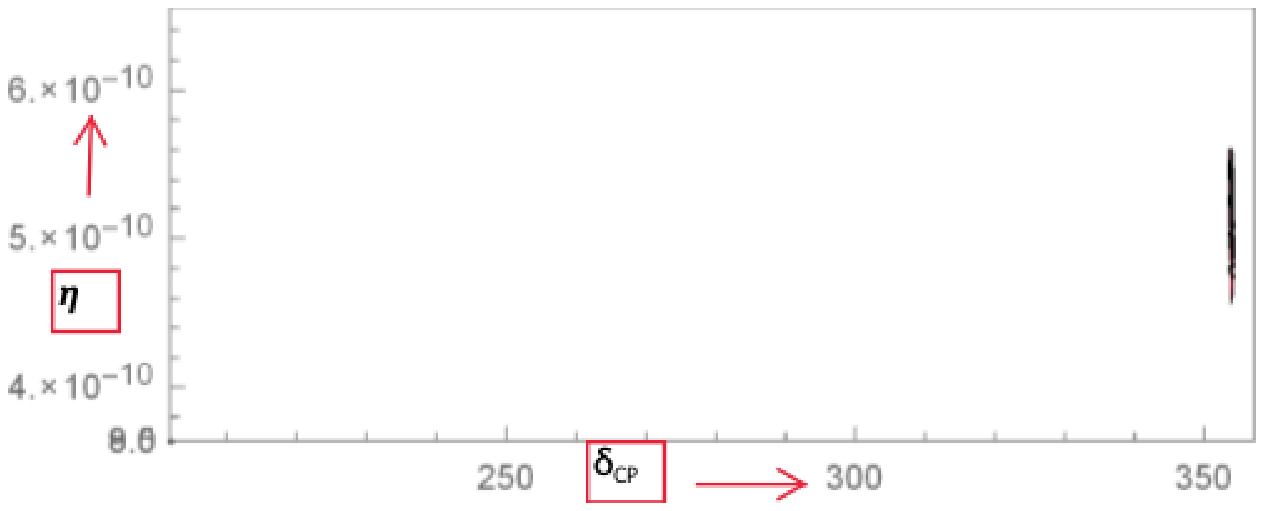}}\end{subfigure}
\begin{subfigure}[]{\includegraphics[height=	6.4cm,width=7.9cm]{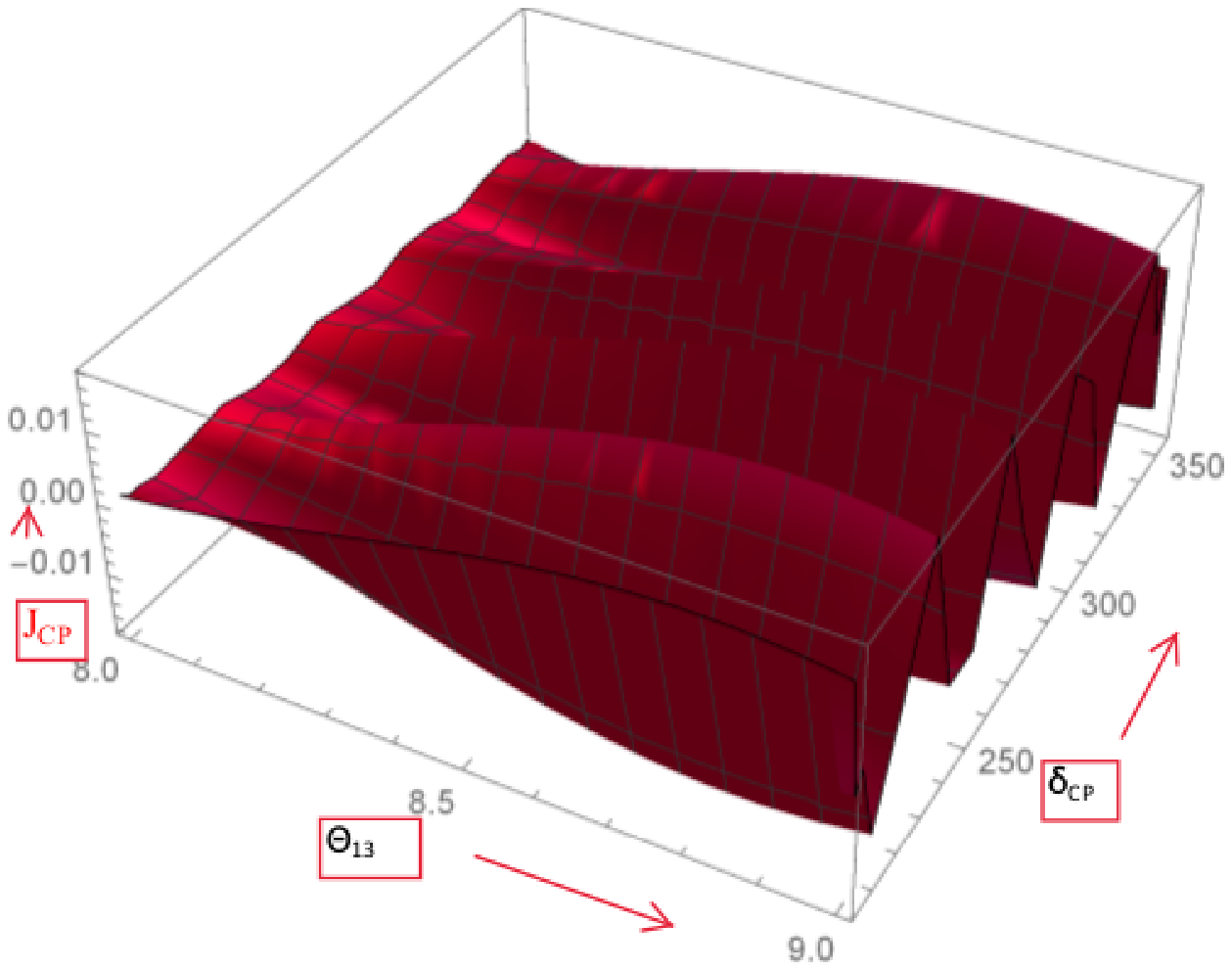}}\end{subfigure}\\

\caption{Predictions in broken $ \mu-\tau $ symmetry model for Inverted ordering. The left panel: predicted favoured values of $ \delta_{CP}$ (in the light of recent ratio of the baryon to photon density bounds, $ 5.8 \times 10^{-10} < \eta < 6.6 \times 10^{-10} $) for lightest $ \nu $ mass $ m_{3} = 0.0657 $ eV, as a result of contribution of type I Seesaw mechanism to neutrino mass matrix. Similarly, The right panel: predicted allowed three dimensional space of $(\delta_{CP}, \theta_{13},J_{CP})$ plane for allowed regions of Jarkslog invariant, $ J_{CP} $ values for for best fit value of $ \theta_{23} = 48.6^{0} $ of $\Delta \chi^{2} = 6.2$ {\color{blue}\cite{pdg}} as a result of contribution of type I Seesaw mechanism to neutrino mass matrix.}} 
\label{fig:1}
\end{figure*}
\end{center}
\begin{center}
\begin{figure*}[htbp]
\includegraphics[height=8cm,width=11cm]{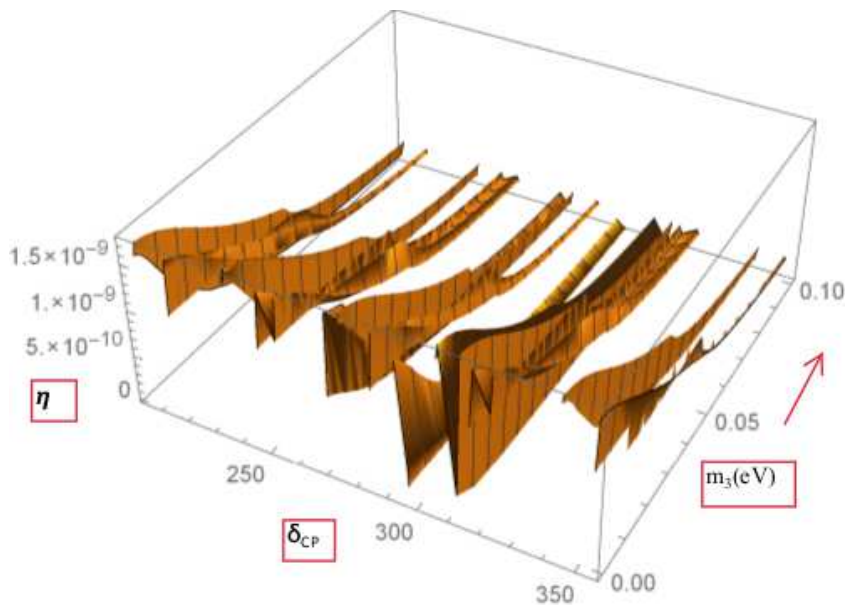}

\caption{Predictions in broken $ \mu-\tau $ symmetry model for Inverted ordering: Three dimensional plot for predicted favoured values of $(m_{3}, \delta_{CP}, \eta )$  plane for best fit values of $ \theta_{13} = 8.49^{0}$ of $\Delta \chi^{2}= 9.5$ w/o SK-ATM {\color{blue}\cite{pdg}} (in the light of recent ratio of the baryon to photon density bounds, $ 5.8 \times 10^{-10} < \eta < 6.6 \times 10^{-10} $) as a result of contribution of type I Seesaw mechanism to neutrino mass matrix.}
\label{fig:1}
\end{figure*}
\end{center}

\begin{center}
\begin{figure*}[htbp]
\centering{
\begin{subfigure}[]{\includegraphics[height=7.4cm,width=7.9cm]{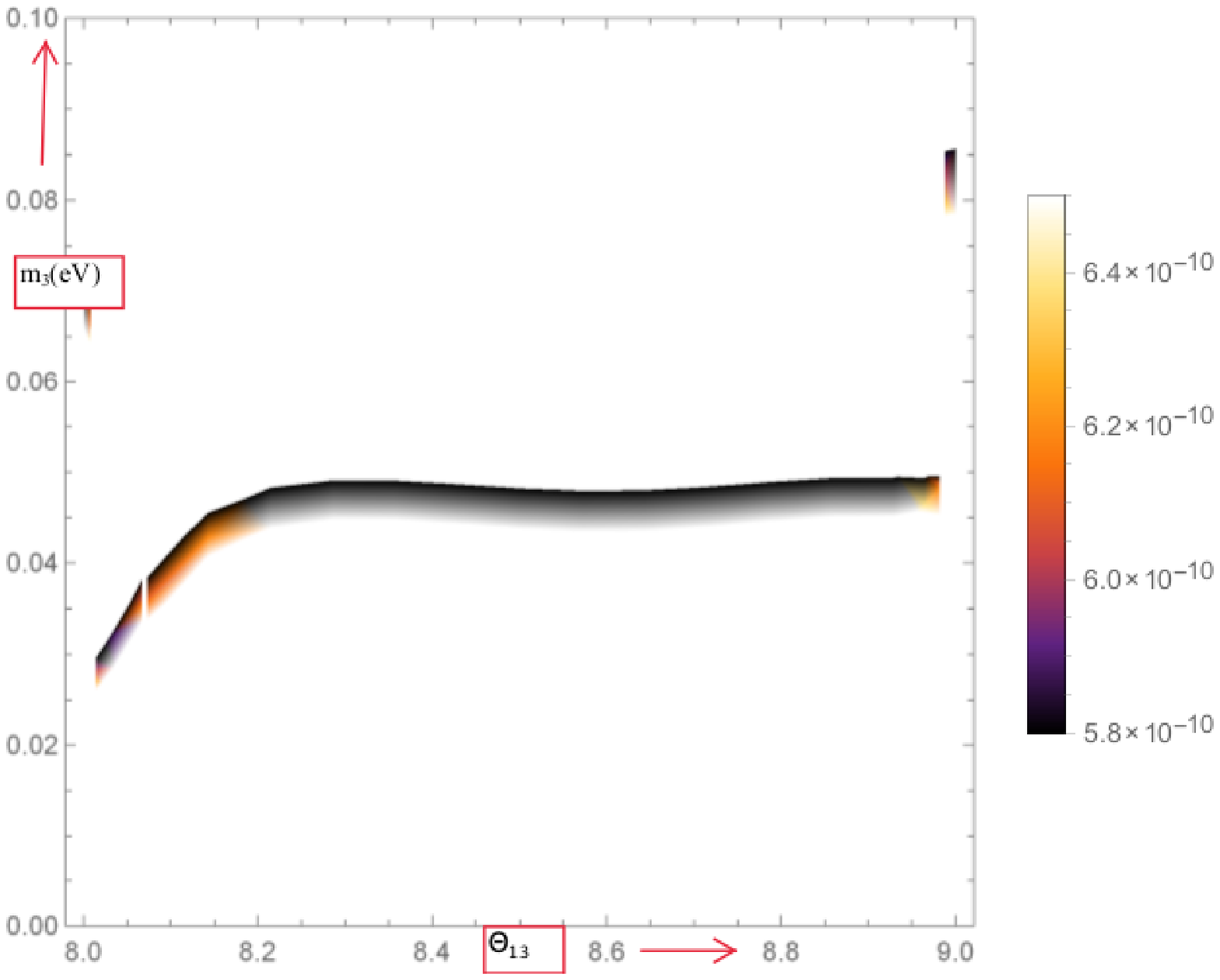}}\end{subfigure}
\begin{subfigure}[]{\includegraphics[height=	7.4cm,width=8.9cm]{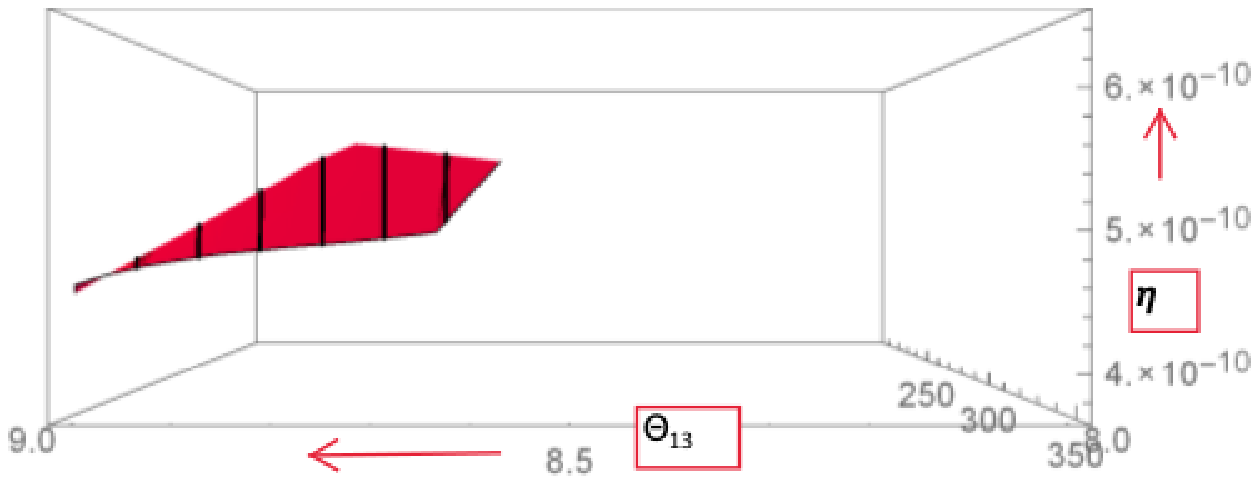}}\end{subfigure}\\

\caption{Predictions in broken $ \mu-\tau $ symmetry model for Inverted ordering. The left panel: density plot of predicted favoured values of $({m_{3}, \theta_{13}})$ plane for best fit values of $ \delta_{CP} = 285^{0}$ of $\Delta \chi^{2}=6.2$ w/o SK-ATM {\color{blue}\cite{pdg}} ( allowed by updated values of correct baryon asymmetry of the Universe) as a result of contribution of type I Seesaw mechanism to neutrino mass matrix. Similarly, The right panel: Three dimensional plot for predicted favoured values of $(\theta_{13}, \delta_{CP}, \eta )$  plane for lightest $ \nu $ mass, $ m_{3} = 0.0657$  (in the light of recent ratio of the baryon to photon density bounds, $ 5.8 \times 10^{-10} < \eta < 6.6 \times 10^{-10} $) as a result of contribution of type I Seesaw mechanism to neutrino mass matrix}}. 
\label{fig:1}
\end{figure*}
\end{center}

\begin{center}
\begin{figure*}[htbp]
\centering{
\begin{subfigure}[]{\includegraphics[height=6.4cm,width=7.9cm]{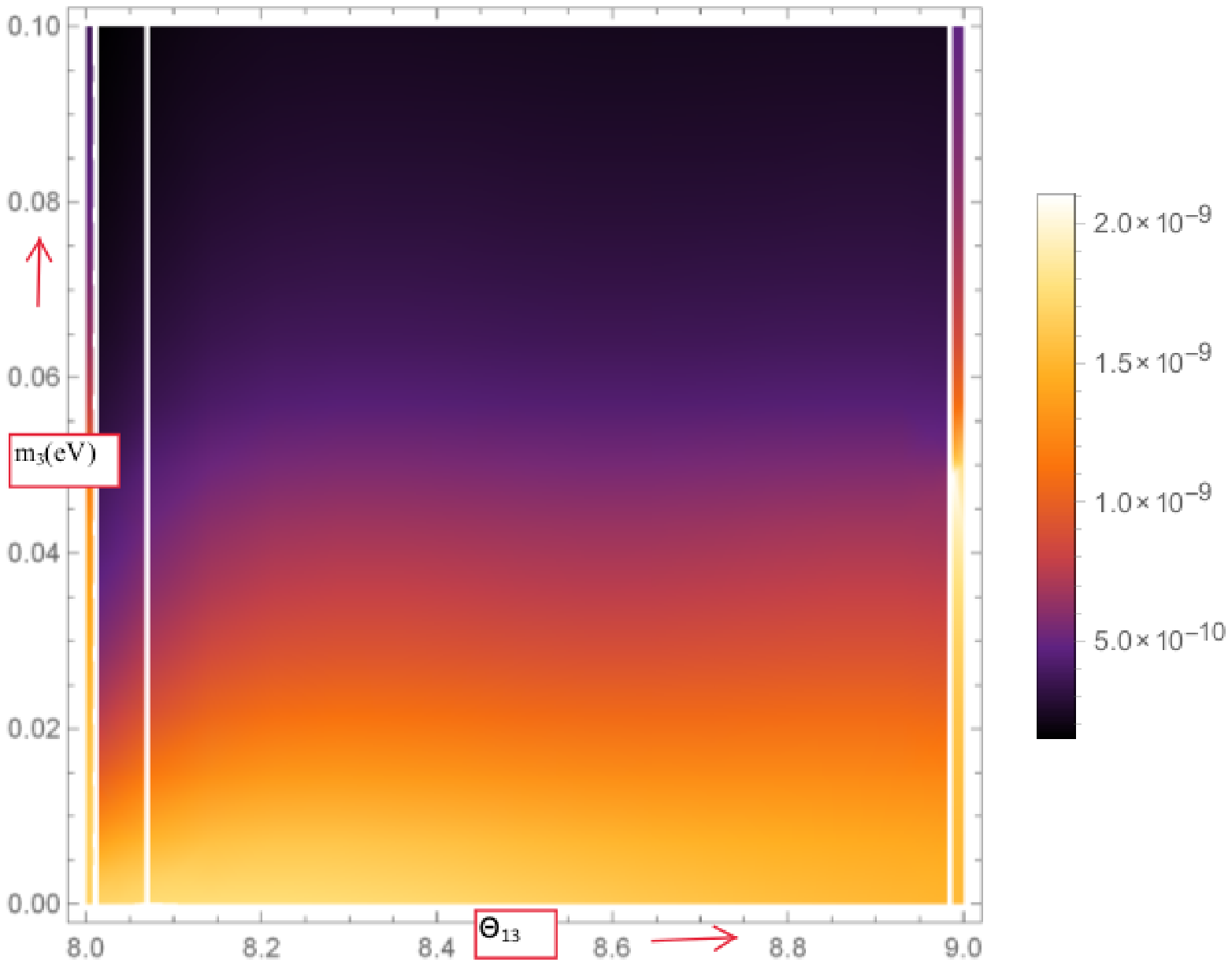}}\end{subfigure}
\begin{subfigure}[]{\includegraphics[height=	6.4cm,width=7.9cm]{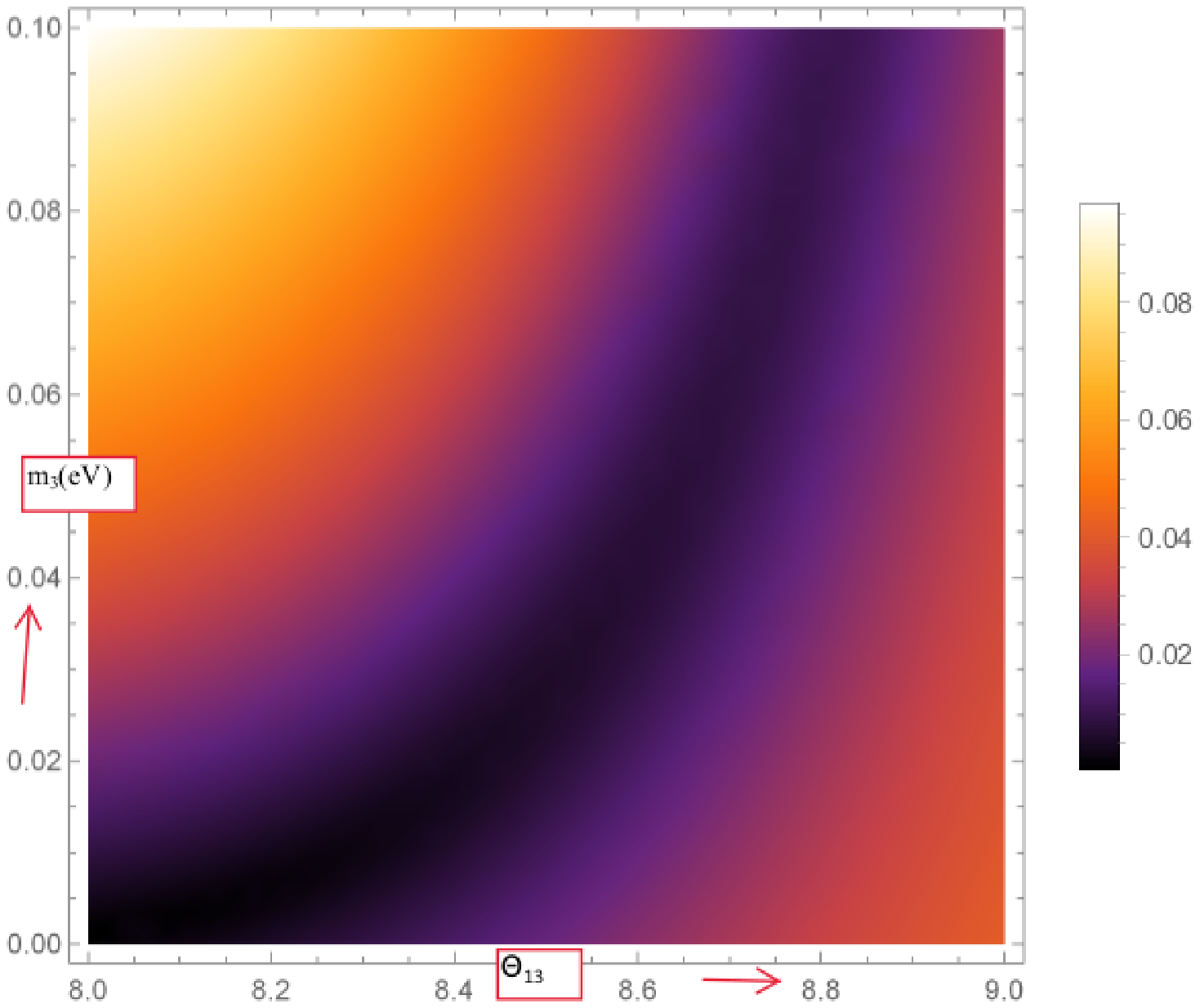}}\end{subfigure}\\

\caption{Predictions in broken $ \mu-\tau $ symmetry model for Inverted ordering. The left panel: density plot of predicted favoured values of $({m_{3}, \theta_{13}})$ plane for best fit values of $ \delta_{CP} = 285^{0}$ of $\Delta \chi^{2}=6.2$ w/o SK-ATM {\color{blue}\cite{pdg}} ( allowed by updated values of correct baryon asymmetry of the Universe) as a result of contribution of type I Seesaw mechanism to neutrino mass matrix. Similarly, The right panel: The predicted two dimensional space of $(m_{3}, \theta_{13})$ for $ m_{ee} $ [eV], $ 0\nu\beta\beta $ decay, for best fit values of $ \delta_{CP} = 285^{0}$ of $\Delta \chi^{2}= 6.2$ w/o SK-ATM {\color{blue}\cite{pdg}}}}. 
\label{fig:1}
\end{figure*}
\end{center}

\begin{center}
\begin{figure*}[htbp]
\centering{

\begin{subfigure}[]{\includegraphics[height=	6.4cm,width=7.9cm]{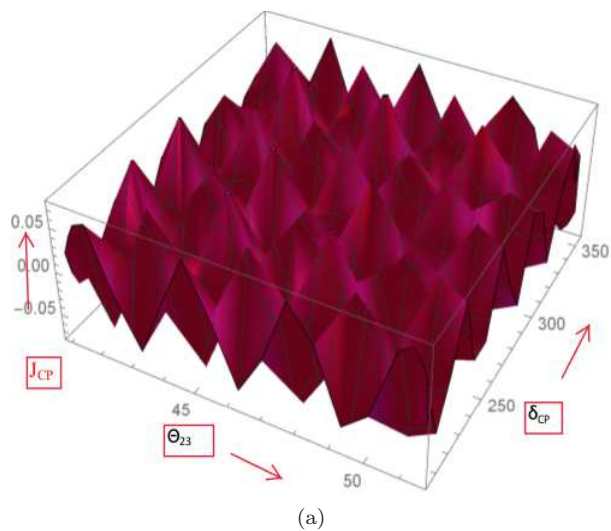}}\end{subfigure}\\

\caption{Predictions in broken $ \mu-\tau $ symmetry model for Inverted ordering: predicted allowed three dimensional space of $(delta_{CP}, \theta_{23},J_{CP})$ plane for allowed regions of Jarkslog invariant, $ J_{CP} $ values for for best fit value of $ \theta_{13} = 8.49 $ of $\Delta \chi^{2} =9.5$ {\color{blue}\cite{pdg}} as a result of contribution of type I Seesaw mechanism to neutrino mass matrix}}. 
\label{fig:1}
\end{figure*}
\end{center}

\begin{center}
\begin{figure*}[htbp]
\centering{
\begin{subfigure}[]{\includegraphics[height=6.4cm,width=7.9cm]{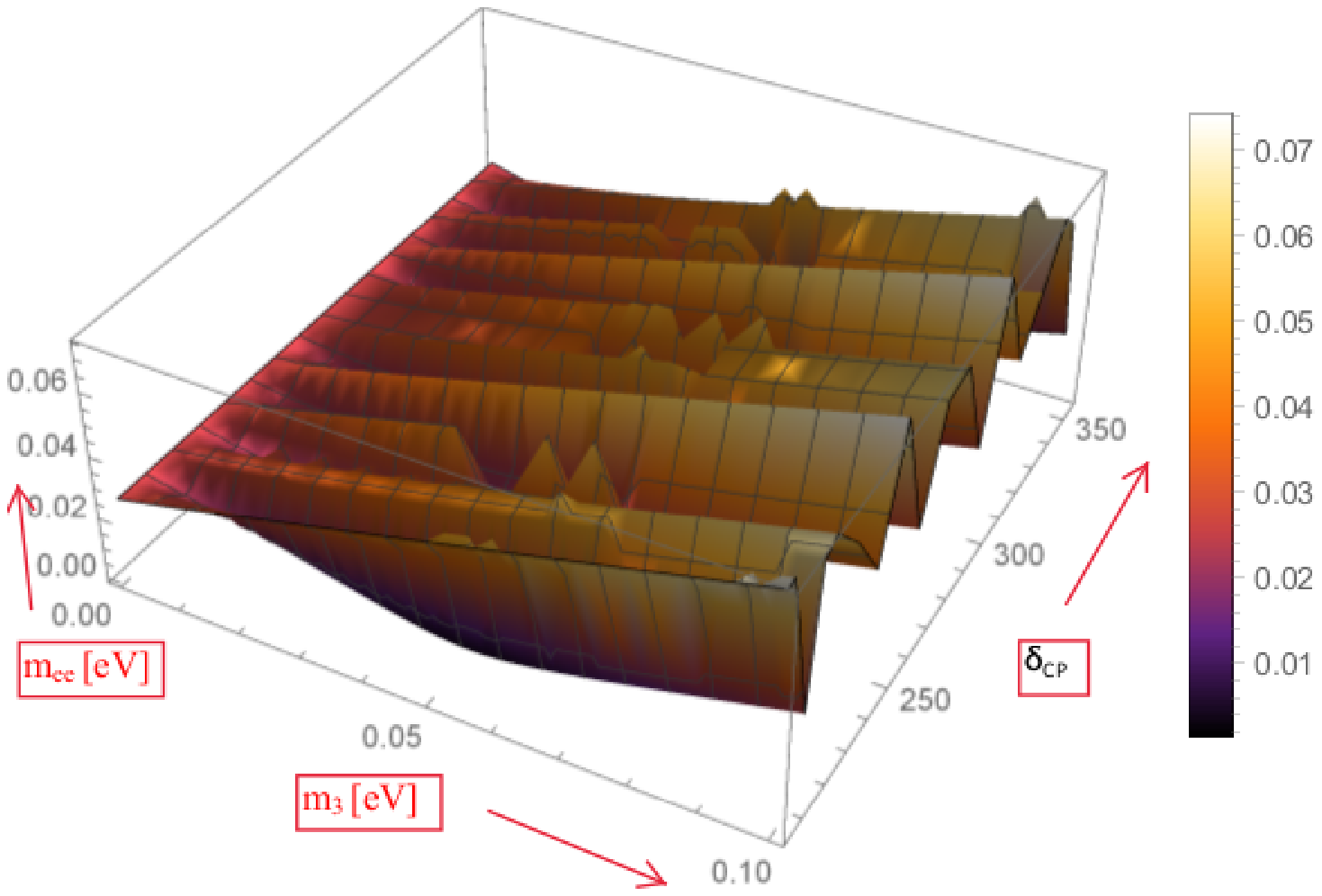}}\end{subfigure}
\begin{subfigure}[]{\includegraphics[height=	6.4cm,width=7.9cm]{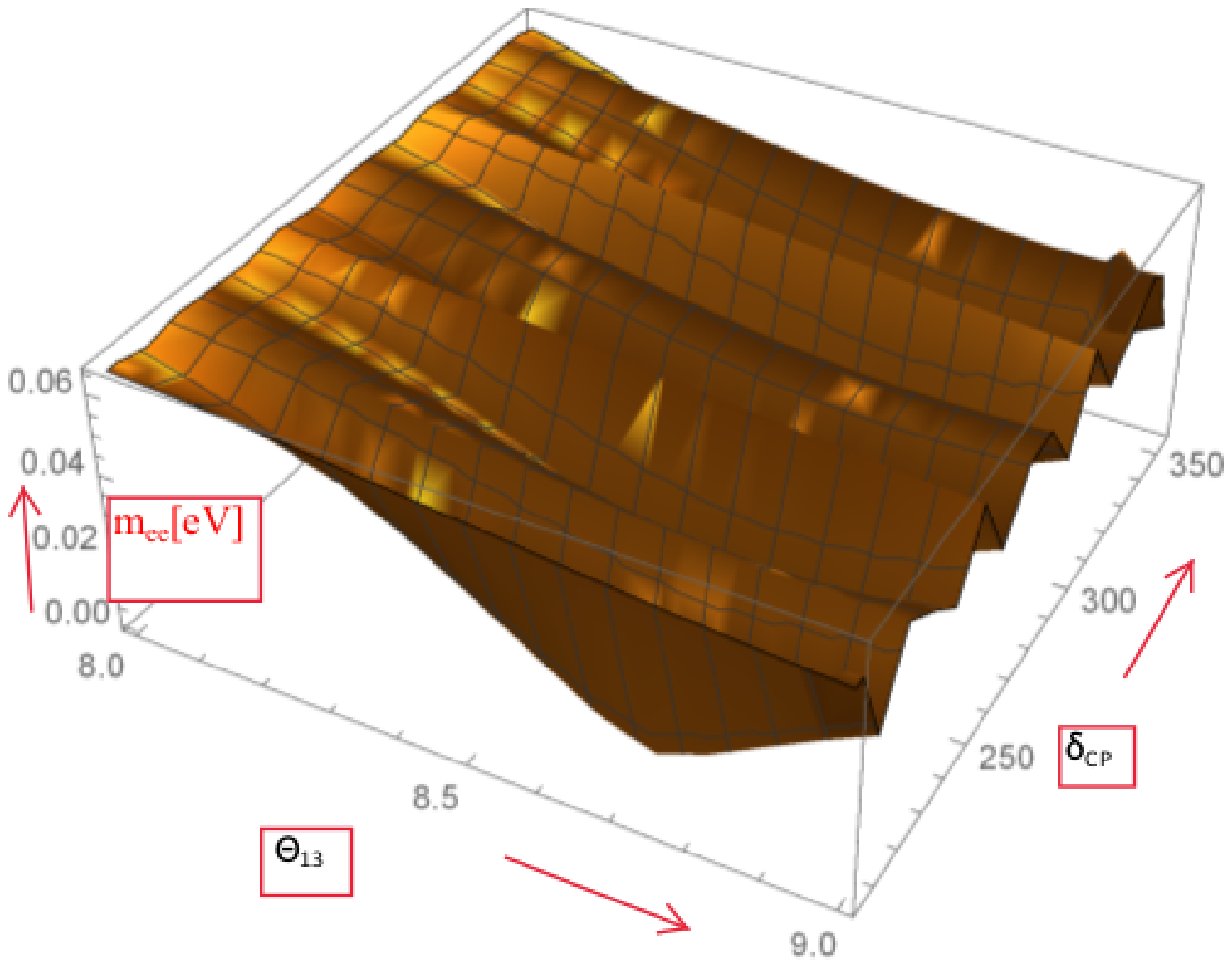}}\end{subfigure}\\

\caption{Predictions in broken $ \mu-\tau $ symmetry model for Inverted ordering. The left panel depicts predicted three
dimensional space of $(m_{ee}, \delta_{CP}, m_{3} )$ for values of $ m_{3} \in [10^{-6}, 0.1]$ eV, $\delta_{CP}$ in the given 3 $ \sigma $range, corresponding to $\Delta \chi^{2} = 6.2$ and best fit values of $ \theta_{13} = 8.49^{0} $ corresponding to $\Delta \chi^{2} = 9.5 $ w/o SK-ATM [27]. The right panel depicts predicted three
dimensional space of $(m_{ee}, \delta_{CP}, \theta_{13} )$ for values of lightest $ \nu $ mass, $ m_{3} = 0.0657 eV$, $\delta_{CP}, \theta_{13}$ in the given 3 $ \sigma $range, corresponding to $\Delta \chi^{2} = 6.2$ and $\Delta \chi^{2} = 9.5 $ w/o SK-ATM [27] }}. 
\label{fig:1}
\end{figure*}
\end{center}

\begin{center}
\begin{figure*}[htbp]
\centering{
\begin{subfigure}[]{\includegraphics[height=6.4cm,width=8.9cm]{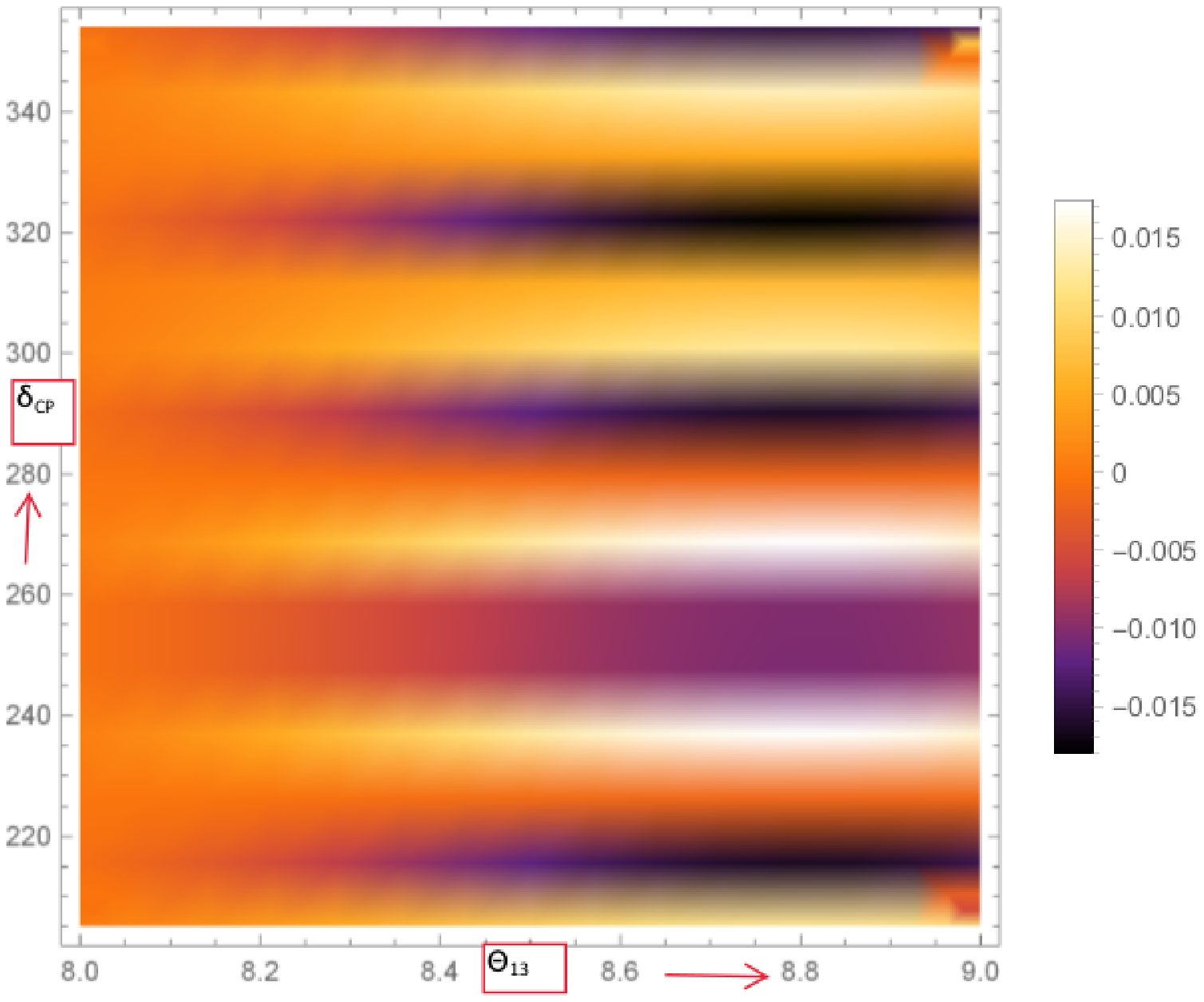}}\end{subfigure}
\\

\caption{Predictions in broken $ \mu-\tau $ symmetry model for Inverted ordering: depicts predicted density plot of $(\delta_{CP}, m_{3} )$ for values of $ m_{3} \in [10^{-6}, 0.1]$ eV, $\delta_{CP}$ in the given 3 $ \sigma $range, corresponding to $\Delta \chi^{2} = 6.2$ and best fit values of $ \theta_{13} = 8.49^{0} $ corresponding to $\Delta \chi^{2} = 9.5 $ w/o SK-ATM [27].}}
\end{figure*}
\end{center}

\section{Results and Conclusion}
In this work, we have used user defined Dirac Neutrino Yukawa couplings for the yukawa interactions associated with the broken $ \mu-\tau $ symmetry model for the generation of          non-zero reactor mixing angle $ \theta_{13} $, leptonic CP phase $ \delta_{CP} $ in type I seesaw Mechanism in the light of leptogenesis and then the transformation of the lepton asymmetry into a baryon asymmetry by non-perturbative $B+L$ violating (sphaleron, sakharov conditions) processes as discussed in {\color{blue}\cite{6}}. A small explicit breaking of $ \mu-\tau  $ symmetry {\color{blue}\cite{kod}}  inherits the property of generating non zero CP violation in $U_{PMNS}$ matrices, $ \delta_{CP} $ phase and results in $ \theta_{13} $ being non zero. Here we consider the type I seesaw as the main donor to neutrino mass. We also take into account both inverted and normal ordering of 
neutrino mass spectrum as well as two different types of lightest neutrino mass $ m_{1}( m_{3}  = 0.07118 eV (0.0657 eV)) $  to visualise the results of hierarchical $ \nu $ mass spectrum. In case of normal ordering of $ \nu $ masses, the dependance of $ \delta_{CP} $ phase on lightest $ \nu $ mass is predicted in figures {\color{red}1,3,4} (in the light of recent ratio of the baryon to photon density bounds, $ 5.8 \times 10^{-10} < \eta < 6.6 \times 10^{-10} $ ). The favoured values of $ \delta_{CP} $ phase is found to lie between $  \delta_{CP}  \in [302^{0}, 304^{0}] $ for best fit values of $ \theta_{13} = 8.41 $ corresponding to $\Delta \chi^{2}=9.5 $ w/o SK-ATM {\color{blue}\cite{pdg}} (in the light of recent ratio of the baryon to photon density bounds $5.8\times 10^{-10} < \eta < 6.6 \times 10^{-10} $) as a result of contribution of type I Seesaw mechanism to neutrino mass matrix. The favoured values of lightest $ \nu $ mass, $ m_{1} $ in this case comes out to be $ \in [0.078,0.1] $ eV. In case of inverted hierarchy the variation of $ \delta_{CP} $ phase is found to be very intense with best fit valuesof $ \theta_{13} = 8.49 $ corresponding to $\Delta \chi^{2}=9.5 $ w/o SK-ATM {\color{blue}\cite{pdg}}. Here only two values of $ \delta_{CP} $ phase is favoured, $ \delta_{CP}  = 220^{0},223^{0},252^{0},268^{0},293^{0},309^{0},345^{0}$ (in the light of recent ratio of the baryon to photon density bounds $5.8\times 10^{-10} < \eta < 6.6 \times 10^{-10} $) as is evident from figure {\color{red}15}. The allowed spectrum of lightest $ \nu $ mass is $ m_{3} $, $ \in [0.02,0.055] $ eV. We also plot the allowed values of $ |m_{ee} |$ eV for neutrinoless double beta decay and the Jarkslog invariant, $ J_{CP} $ in figures {\color{red}5,6,7,8,9,10,11,12,13} for normal ordering of $ \nu $ masses. Prediction of future leptonic CP violation experiments should be able to rule out or take into account some of the results discussed in this work. If we abide by the best fit values of leptonic CP phase $ \delta_{CP} = \frac{3}{2}\pi$ discussed in the literature {\color{blue}\cite{r}} then our scenario, $  \delta_{CP}  \in 268^{0}$  for inverted ordering of $ \nu $ masses favours the value of  $ \delta_{CP} = \frac{3}{2}\pi$ for T2K neutrino and antineutrino appearance results. We show the variation of baryon asymmetry with leptonic $ \delta_{CP} $phase in table I. 
\begin{table}[htb]
\renewcommand{\arraystretch}{1.5}
\begin{center}
\begin{tabular}{|c|c|}
\hline 
 $ \mu-\tau $ broken symmetry model & Calculated leptonic CP phase $\delta_{CP} $\\ 
\hline 
 $m_{1}  \in [0.078,0.1]$ eV (NH)   & $ \delta_{CP} \in [302^{0},304^{0}]$\\
\hline
 $m_{3}  \in [0.02,0.055]$ eV (IH)   & $ \delta_{CP}  = 220^{0},223^{0},252^{0},268^{0},293^{0},309^{0},345^{0}$\\

\hline 
\end{tabular}
\end{center}
\caption{Values of $ \delta_{CP} $ phase giving correct updated values of baryon asymmetry.}
\end{table}
We also plot the allowed values of $ |m_{ee} |$ eV for neutrinoless double beta decay and the Jarkslog invariant, $ J_{CP} $ in figures {\color{red}14,17,18,19,20} for inverted ordering of $ \nu $ masses. Future LBL experiments, will hunt for the leptonic CP phase and potentially will measure it with precision. Neutrinoless double beta decay will indicate towards Majorana CPV phase. The fundamental mysteries in the Universe are about the findings of the nature of the massive neutrinos-Dirac or Majorana. This may be sorted out by the experiments like GERDA, CUORE, KamLAND-Zen, EXO, LEGEND, nEXO etc. Determination of the status of leptonic CP asymmetry (T2K, NO$ \nu $A, T2HK, DUNE), determination of type of neutrino mass ordering (T2K+NO$ \nu $A, JUNO, PINGU, ORCA, T2HKK, DUNE)  and determination of the order of absolute neutrino mass scale (KATRIN, Cosmology) are few of the most challenging tasks today. The ideas presented in this work may definitely will rule in or rule out some of the favoured space in few of the above experiments.
 
\section{Acknowledgement}
I would like to thank my supervisor Prof. Kalpana Bora for useful discussion on this topic.

\end{document}